\documentclass[aip,amsmath,amssymb,reprint]{revtex4-1}

\usepackage{graphicx}
\usepackage{dcolumn}
\usepackage{bm}

\usepackage[utf8]{inputenc}
\usepackage[T1]{fontenc}
\usepackage{mathptmx}
\usepackage{soul}

\newcommand{\p}{\partial} 
\newcommand{\vx}{\vec{x}}
\newcommand{\vy}{\vec{y}}
\newcommand{\vz}{\vec{z}}

\newcommand{\vv}{\vec{v}}

\newcommand{\vp}{\vec{p}}
\newcommand{\vq}{\vec{q}}

\newcommand{\vj}{\vec{j}}

\usepackage{xcolor}

\begin{document}


\title[FRG approach to scalar turbulence]{Spatio-temporal correlation functions in scalar turbulence from functional renormalization group}

\author{Carlo Pagani}
\author{L\'eonie Canet}%
\altaffiliation{
Institut Universitaire de France, 1 rue Descartes, 75000 Paris, France.
}
\affiliation{ 
Universit{\'e} Grenoble Alpes, Centre National de la Recherche Scientifique, 
Laboratoire de Physique et Mod{\'e}lisation des Milieux Condens{\'e}s, 38000 Grenoble, France.
}%


\begin{abstract}
We provide the  leading behavior at large wavenumbers of the two-point correlation function
 of a scalar field passively advected by a turbulent flow.
We first consider the Kraichnan model, in which the turbulent carrier flow is modeled
 by a stochastic vector field with a Gaussian distribution,
 and then a  scalar
  advected by a homogeneous and isotropic turbulent flow described by the Navier-Stokes equation,  under the assumption
that the scalar is passive, {\it i.e.}, that it does not affect the carrier flow. We show that at large wavenumbers, the two-point correlation function of the scalar in the Kraichnan model
    decays as an exponential in the time delay, in both the inertial and dissipation ranges.
 We establish the expression, both from a perturbative and from a nonperturbative calculation, of the prefactor, which is found to be always proportional to $k^2$.
  For a real scalar, the decay is Gaussian in $t$ at small time delays, and it crosses over to an exponential
  only at large $t$. The assumption of delta-correlation in time of the stochastic velocity field in the Kraichnan model  hence  significantly alters the statistical temporal behavior of the scalar at small times.

\end{abstract}

\maketitle

\section{Introduction}

The advection of scalar fields by external flows enters in a wide
range of phenomena, such as the fluctuations of temperature or salinity
 in the ocean, or the dispersion of pollutants in the atmosphere.
  When the carrier
   flow is turbulent, understanding 
  the statistical properties and the mixing  of the scalar
   become a difficult and fundamental issue.
  A simplifying assumption is to consider as passive
   the advected scalar field, which means that it has a negligible backreaction onto the
 carrier flow, such that the dynamics of the latter is unaffected.
 This assumption is realistic in particle-ladden flows only at sufficiently small concentration.
  However, even in this simpler case, the complete
   characterization of the properties of the scalar is still challenging, and we focus
    on this case in this work.
 
 For a high-Reynolds number carrier flow, 
 one can distinguish several regimes for the scalar
  depending on the Schmidt number, which is the ratio 
  of the fluid viscosity to the scalar diffusivity.
  We refer to Ref.~\onlinecite{Sreenivasan18175} for an overview.
 In this work, we  focus on the inertial-convective range, which corresponds
  to Schmidt numbers of order one, and which spans scales between
   the energy injection scale of the scalar and the scale at which energy is dissipated, 
   for typical scales of the velocity of the turbulent fluid in the inertial range. 
   The spectrum of the scalar field in the inertial-convective range
 was established by Obukhov and Corrsin \cite{Obukhov49,Corrsin51}.
 Other pioneering studies are due to Yaglom, Batchelor, and Kraichnan
\cite{Yaglom49,batchelor_1959,Kraichnan_PoF_68}. Barring possible
corrections due to intermittency, the scalar spectrum in this
range is determined via the energy cascade picture and shows a power-law decay with the
same exponent $-5/3$ in three dimensions as the spectrum of the turbulent fluid carrying it. 
  
 The spectrum is related to the two-point correlation function of the scalar field 
 at equal times. Characterizing its behavior at different times 
  is also fundamental to understand the mixing properties of the scalar.
  In particular, determining the temporal dependence of the Eulerian correlations of the scalar
  is a difficult task and few is known so far on their properties. 
    In this work, we use the framework of the
   functional renormalization group (FRG) (see Ref.~\onlinecite{Delamotte:2007pf} for an introduction and Ref.~\onlinecite{Dupuis:2020fhh} for a recent review) to make progress on this issue.
    The FRG approach was recently employed to study  Navier-Stokes (NS) turbulence, where it proved
     to be fruitful. By exploiting the symmetries of the field
theoretical formulation of the stochastically forced Navier-Stokes equation, it 
   yielded predictions for the general form of the spatio-temporal dependence
 of any $n$-point correlation function of the turbulent velocity field in the limit of large wavenumbers  \cite{Canet:2016nmg,Tarpin:2017uzn}. 
 These predictions were accurately confirmed
     by Direct Numerical Simulations \cite{Canet:2016nmg,Gorbunova:2020uxl,Gorbunova:2021a}.
 The FRG approach was also extended to study the direct cascade in 2D turbulence \cite{Tarpin:2018yvs}. 

  In this work, we follow a similar strategy for the passive scalar, in order to construct
 an approximation scheme which is not based on a
 small-coupling expansion, and thus offers a nonperturbative approach.
 Exploiting the symmetries of the field theory of the advected scalar,
 we provide the general expression of the temporal dependence of the two-point Eulerian correlations
  of the scalar field in the limit of large wavenumbers.
  These expressions are obtained both for 
  a scalar advected by a turbulent NS flow, and for a scalar advected by a ``synthetic'' stochastic velocity field, as prescribed in the  Kraichnan model
\cite{Kraichnan_PoF_68,Kraichnan94prl}. In the latter case,
 the velocity statistics is Gaussian and the associated covariance is white-in-time. Although these
  delta-correlations yield simpler explicit expressions, they  have a significant impact on the
  temporal behavior of the scalar field at small time. Indeed, we show that for a scalar in a turbulent NS flow,
    the two-point correlation function decays at small $t$ as a Gaussian in the variable $tk$ where $t$ is the time delay
   and $k$ the wavenumber, and as an exponential at large $t$. 
   The scalar field in the inertial-convective regime thus inherits in this case  temporal correlations which are 
   very similar to the ones of the turbulent  fluid (reported {\it eg.} in Ref.~\onlinecite{Tarpin:2018yvs}). 
  For the Kraichnan scalar, the two-point correlations always decay exponentially, the Gaussian small-time behavior is completely washed out by the instantaneous decorrelation of the synthetic velocity field.

The paper is organized as follows.
In Sec.~\ref{sec:Kraichnan-model}, we introduce the Kraichnan model, 
the associated field theory, and we expound its symmetries and the related Ward identities.
In Sec.~\ref{sec:preliminaries-and-non-renormalization-th},
we establish some preliminary results for this model that will be used to derive a closed
flow equation for the correlation functions of the Kraichnan scalar in Sec.~\ref{sec:closed-FRGE-Kraichnan}.
In Sec.~\ref{sec:closed-FRGE-for-advected-scalars}, we generalize
the derivation and results to the case of scalar fields advected by turbulent NS flows.
We summarize our findings in Sec.~\ref{sec:summary}.
In Appendix \ref{app:Yaglom_rel}, we discuss the Yaglom relation 
from a field theoretical viewpoint, and in Appendix \ref{app:dissipative},
we establish the expression of the spectrum in the near-dissipative range from
 the Dyson equation.

\section{Field theory and symmetries of the Kraichnan model} \label{sec:Kraichnan-model}

The Kraichnan model  was proposed as
 a simplified model of scalar turbulence where the advecting velocity field 
 is taken as a random vector field with  Gaussian statistics to replace 
  the NS fluctuating turbulent velocity field \cite{Kraichnan_PoF_68}.
The key simplifying feature of this model is that the velocity covariance
 is white in time,  $\langle v^{i}\left(t,\vx\right)v^{j}\left(t^{\prime},\vx\right)\rangle\propto\delta\left(t-t^{\prime}\right)$.
 This model has been widely studied since many aspects can be investigated analytically. 
 In particular, several techniques have been employed to compute  
intermittency corrections via suitable expansions \cite{ChertkovETal1995,Gawedzki:1995zz,ChertkovFalkovich1996,Bernard:1996um,BernardGK1998}.
In the perturbative regime, intermittency corrections were computed also
via various field theoretical approaches \cite{PhysRevE.58.1823,PhysRevE.63.025303,PhysRevE.58.7381,Kupiainen:2006em}, including the functional renormalization group \cite{Pagani:2015hna}.

In this section, we present the path integral formulation of the Kraichnan
model,  introduce the FRG formalism, and then discuss the symmetries of the Kraichnan field theory and
the associated Ward identities. We shall focus on incompressible flows 
and take the velocity field to be divergenceless.

\subsection{The Kraichnan model}

The Kraichnan model describes the advection and diffusion of a scalar
field $\theta\left(t,\vx\right)$ following the dynamics
\begin{eqnarray}
\partial_{t}\theta\left(t,\vx\right)+v^{i}\left(t,\vx\right)\partial_{i}\theta\left(t,\vx\right)-\frac{\kappa}{2}\partial^{2}\theta\left(t,\vx\right) & = & f\left(t,\vx\right)\,,\label{eq:Kraichnan-model_SDE}
\end{eqnarray}
where $\kappa$ is the molecular diffusivity and $f\left(t,\vx\right)$
is a stochastic forcing characterized by  Gaussian statistics with
covariance $\langle f\left(t,\vx\right)f\left(t^{\prime},\vy\right)\rangle=\delta\left(t-t^{\prime}\right)M\left(\frac{|\vx-\vy|}{L}\right)$. In this model,
the velocity $v^{i}\left(t,\vx\right)$ is a stochastic vector field chosen with 
a Gaussian distribution. It has zero average and  the following covariance:
\begin{eqnarray}
\Bigr\langle v^{i}\left(t,\vx\right)v^{j}\left(t^{\prime},\vy\right)\Bigr\rangle & = & \delta\left(t-t^{\prime}\right)D_{0}\int_{\vp} \frac{e^{i \vp\cdot \left(\vx-\vy\right)} \,P_{ij}(\vp)}{\left(p^{2}+m^{2}\right)^{\frac{d}{2}+\frac{\varepsilon}{2}}} 
\,,\label{eq:vel-2pt-correlation-function}
\end{eqnarray}
with $P_{ij}(\vp)=\delta_{ij}-\frac{p_{i}p_{j}}{p^{2}}$  the transverse
projector. The mass $m$ in Eq.~(\ref{eq:vel-2pt-correlation-function})
prevents any infrared divergence in the velocity covariance. 
Note that we use for the integrals the
 shorthand notation 
\begin{equation*}
\int_{t\vx}  =  \int dt\int d^{d}\vx\quad,\quad \quad \int_{\omega \vq}  =  \int\frac{d\omega}{2\pi}\int\frac{d^{d}\vq}{\left(2\pi\right)^{d}}\,
\end{equation*}
and for the Fourier transforms the convention 
\begin{eqnarray}
g\left(\omega,\vp\right) & \equiv & \int_{t\vx}e^{i\omega t-i\vp\cdot\vx}g\left(t,\vx\right),\nonumber\\
 g\left(t,\vx\right) & \equiv & \int_{\omega \vp}e^{-i\omega t+i\vp\cdot \vx}g\left(\omega,\vp\right)\,.
\label{eq:Fourier}
\end{eqnarray}
In the limit $\varepsilon\rightarrow0$, the coinciding-point
covariance (\ref{eq:vel-2pt-correlation-function}) diverges logarithmically.
Formally, a finite limit can be obtained by defining $D_{0}\equiv D_{0}^{\prime}\varepsilon$.
We shall use either $D_{0}$ or $D_{0}^{\prime}$, keeping in mind that they are just fixed numbers for a
given value of $\varepsilon$.

The stochastic differential equation (SDE) Eq.~(\ref{eq:Kraichnan-model_SDE})
is a multiplicative one and is interpreted
in the Stratonovich sense. It can also be rewritten  
in the Ito sense, but we shall work here with the Stratonovich convention unless otherwise stated.
The Martin-Siggia-Rose-Janssen-de Dominicis (MSRJD) formalism provides
a well established procedure to map a SDE
  to an equivalent path integral formulation \cite{Martin73,Janssen76,Dominicis76,Dominicis79}.
The MSRJD approach allows one to compute an arbitrary correlation
 function by introducing the appropriate generating functionals. 
 A key advantage of employing the MSRJD formalism is that
 it allows one to compute Eulerian correlation functions at different 
 times in a simple way.

 The generating functional for the  Kraichnan model is obtained as
 \begin{equation}
  {Z} = \int {\cal D}\theta{\cal D}\bar{\theta} {\cal D}c{\cal D}\bar{c}{\cal D}\vv e^{-S_{{\rm Kr}}  + \int_{t,\vx} \phi {J}}
 \end{equation}
with the following action:
\begin{eqnarray}
S_{{\rm Kr}} & = & \int_{t\vx}\,\bar{\theta}\left(t,\vx\right)\Biggr\{\partial_{t}\theta\left(t,\vx\right)+v^{i}\left(t,\vx\right)\partial_{i}\theta\left(t,\vx\right)-\frac{\kappa}{2}\partial^{2}\theta\left(t\vx\right)\Biggr\}\nonumber \\
 &  & -\frac{1}{2}\int_{t\vx \vy} \bar{\theta}\left(t,\vx\right)M\left(\frac{\vx-\vy}{L}\right)\bar{\theta}\left(t,\vy\right)\nonumber \\
 &  & +\int_{t\vx}\,\bar{c}\left(t,\vx\right)\Biggr\{\partial_{t}c\left(t,\vx\right)+v^{i}\left(t,\vx\right)\partial_{i}c\left(t,\vx\right)-\frac{\kappa}{2}\partial^{2}c\left(t,\vx\right)\Biggr\} \nonumber \\
 &  & +\frac{1}{2}\int_{t\vx}\,v^{i}\left(t,\vx\right)\frac{\left(-\partial^{2}+m^{2}\right)^{\frac{d}{2}+\frac{\varepsilon}{2}}}{D_{0}}v^{i}\left(t,\vx\right) \,.\label{eq:Kraichnan-model-action}
\end{eqnarray}
Several comments regarding this action  are in order.
The field $\bar{\theta}$ is usually called the response field
since it is related to  response functions. The quadratic term in
$\bar{\theta}$ originates from integrating out the stochastic forcing
$f$ of the scalar: the function $M\left(\frac{\vx-\vy}{L}\right)$ is essentially the
forcing covariance and $L$ represents the large length scale at which
energy is injected to stir the scalar. The third line in (\ref{eq:Kraichnan-model-action})
contains the Grassmannian odd fields $c$ and $\bar{c}$ which are
introduced to express the functional determinant involved in the procedure via an action and are referred
to as ghosts. This determinant is written in the Stratonovich
convention. The source term ${J}$ gathers the different sources, 
respectively $J_{\theta},\bar{J}_\theta,\vj,\eta,\bar{\eta}$, for the fields $\theta,\bar{\theta},\vv,c,\bar{c}$
 gathered in the field multiplet $\phi$.

\subsection{The functional renormalization group formalism} \label{subsec:FRG-formalism}

The renormalization group has been instrumental for understanding
the scaling properties of physical systems at criticality. Also away
from equilibrium the RG constitutes a crucial conceptual and calculational
framework to investigate several phenomena. As far as turbulence is
concerned, the renormalization group approach has been employed in
many studies \cite{Dominicis79,Fournier83,Yakhot1986,Canuto96,Adzhemyan03,Adzhemyan1999}. 
However, perturbative techniques have often been
hampered by the lack of a small expansion parameter. 

The functional renormalization group implements the Wilsonian renormalization
program at a functional level by adding to the microscopic action
a term $\Delta S_{k}$ which introduces a wavenumber scale $k$ and suppresses
the integration over low wavenumber modes $|\vp|\lesssim k$ 
\cite{Wetterich:1992yh,Reuter:1993kw,Reuter:1994sg,Ellwanger:1993mw,Morris:1993qb}.
The fluctuations are hence progressively averaged thereby building up the effective
 theory at scale $k$.
The  modified effective action in the presence of this scale-dependent term
 is called the effective average action
(EAA) and its $k$-dependence is governed by an exact equation, which
can be solved by implementing some approximation scheme. We refer
to  Ref.~\onlinecite{Delamotte:2007pf} for a pedagogical introduction  and to  Ref.~\onlinecite{Dupuis:2020fhh}
 for a recent review. 

 To achieve the separation of fluctuation modes, the term $\Delta S_{k}$ is chosen quadratic in the
fields:
\begin{eqnarray*}
\Delta S_{k}\left[\phi\right] & \equiv & \int_{t\vx}\frac{1}{2}\phi R_{k}\left(-\partial^{2}\right)\phi\,,
\end{eqnarray*}
where $R_{k}\left(-\partial^{2}\right)$ is a suitable cutoff kernel.
It is required to be large for modes with wavenumber $|\vp| \lesssim k$ 
and to vanish for modes with   $|\vp| \gtrsim k$ such that only these modes are integrated out:
\begin{equation}
 R_k(p^2) \stackrel{|\vp|\lesssim k}{\sim} k^2\quad\quad\quad R_k(p^2) \stackrel{|\vp|\gtrsim k}{\longrightarrow}0
\label{eq:cutoff}
\end{equation}
The modified functional integral takes the form
\begin{eqnarray*}
Z_k = e^{W_{k}\left[J\right]} & \equiv & \int{\cal D}\phi\,e^{-S\left[\phi\right]-\Delta S_{k}\left[\phi\right]+\int_{t\vx} {J}\phi}\,,
\end{eqnarray*}
where $W_{k}\left[J\right]$ is the (modified) generating functional
of the connected correlation functions.
The EAA is defined by the Legendre transform of $W_{k}\left[J\right]$
and is denoted by $\Gamma_{k}\left[\varphi\right]$:
\begin{eqnarray}
\Gamma_{k}\left[\varphi\right]+\Delta S_{k}\left[\varphi\right] & = & \int_{t\vx} J\varphi-W_{k}\left[\varphi\right]\,,
\label{eq:Legendre}
\end{eqnarray}
where $\varphi$ is the average field. The scale dependence of the
EAA is governed by an exact RG flow equation that takes the following general
form:
\begin{eqnarray}
\partial_{s}\Gamma_{k}\left[\varphi\right] & = & \frac{1}{2}\mbox{Tr}\left[\left(\Gamma_{k}^{\left(2\right)}\left[\varphi\right]+R_{k}\right)^{-1}\partial_{s}R_{k} \right]\,,
\label{eq:FRGE-generic_EAA}
\end{eqnarray}
where $s\equiv\log (k/\Lambda)$ with $\Lambda$ a UV scale.
At this scale, one can show
  that $\Gamma_\Lambda$ identifies with the bare action, since no fluctuation is yet incorporated.
   When $k\to 0$, the regulator is removed and one obtains the actual properties of the model, when all fluctuations
   have been integrated over. Eq.~(\ref{eq:FRGE-generic_EAA}) provides the exact interpolation between these two scales.

After this general introduction, let us discuss the application of
the FRG formalism to the Kraichnan model. In particular, we follow
the approach discussed in  Ref.~\onlinecite{Canet:2014dta}, where the scale
$k$ is identified with the forcing scale of the scalar, i.e. $k\equiv L^{-1}$, which is now running,
 and $M\to M_k$ in Eq.~(\ref{eq:Kraichnan-model-action}).
Indeed, since the forcing covariance is peaked at large
length scales and vanishes at small length scales, it satisfies the requirements (\ref{eq:cutoff})  and
can therefore be interpreted as a cutoff. Besides this ``effective forcing'' cutoff,
 we also introduce  cutoff kernels for the other fields. The  generalized
cutoff action is defined by
\begin{eqnarray}
\Delta S_{k}\left[\phi\right] & \equiv & \int_{t\vx}\,\Biggr\{\frac{\kappa}{2}\bar{\theta}R_{k}\left(-\partial^{2}\right)\theta-\frac{1}{2}\bar{\theta}M_{k}\left(-\partial^{2}\right)\bar{\theta}\nonumber \\
 &  & +\frac{1}{2D_{0}}v^{i}R_{k,vv}\left(-\partial^{2}\right)v^{i}+\frac{\kappa}{2}\bar{c}R_{k}\left(-\partial^{2}\right)c\Biggr\}\,.\label{eq:cutoff-action-Kraichnan}
\end{eqnarray}
With this addition the total action finally reads
\begin{eqnarray}
&& S = \int_{t\vx}\,\bar{\theta}\left(t,\vx\right)\Biggr\{\partial_{t}\theta\left(t,\vx\right)+v^{i}\left(t,\vx\right)\partial_{i}\theta\left(t,\vx\right)-\frac{\kappa}{2}\partial^{2}\theta\left(t,\vx\right)\Biggr\}\nonumber \\
 &  & + \int_{t\vx}\,v^{i}\left(t,\vx\right)\frac{\left(-\partial^{2}+m^{2}\right)^{\frac{d}{2}+\frac{\varepsilon}{2}}}{2 D_{0}}v^{i}\left(t,\vx\right)\nonumber \\
 &  & +\int_{t\vx}\,\bar{c}\left(t,\vx\right)\Biggr\{\partial_{t}c\left(t,\vx\right)+v^{i}\left(t,\vx\right)\partial_{i}c\left(t,\vx\right)-\frac{\kappa}{2}\partial^{2}c\left(t,\vx\right)\Biggr\} \label{eq:S_final_total} \\
 &  & +\int_{t\vx}\,\Biggr\{\frac{\kappa}{2}\bar{\theta}\left(t,\vx\right)R_{k}\left(-\partial^{2}\right)\theta\left(t,\vx\right)-\frac{1}{2}\bar{\theta}\left(t,\vx\right)M_{k}\left(-\partial^{2}\right)\bar{\theta}\left(t,\vx\right)\nonumber \\
 &  & +v^{i}\left(t,\vx\right) \frac{R_{k,vv}\left(-\partial^{2}\right)}{2D_{0}} v^{i}\left(t,\vx\right)+\frac{\kappa}{2}\bar{c}\left(t,\vx\right)R_{k}\left(-\partial^{2}\right)c\left(t,\vx\right)\Biggr\}\,, \nonumber
\end{eqnarray}
where the last two lines contain the cutoff action which incorporates the scalar forcing term.

\subsection{Vertex functions and notation}

The EEA $\Gamma_k$ introduced in section \ref{subsec:FRG-formalism} is the generating functional of the one-particle-irreducible (1-PI) correlation functions,
 also called vertices, which are often introduced in field theoretical treatments.
 The set of $n$-point connected correlation functions, denoted generically $W_k^{(n)}$, is equivalent
  to the set of $n$-point 1-PI correlation functions, denoted $\Gamma_k^{(n)}$, in the sense that one can be constructed from
   the other. Both types of correlation functions are used in the following.	
Let us introduce some notation, in order to specify the fields entering these functions.
By taking $n$ functional derivatives of the EEA, one obtains the $n$-point corresponding vertex function
\begin{eqnarray*}
&& \frac{\delta^{n}\Gamma_{k}}{\delta\varphi_{1}\left(t_{1},\vx_{1}\right)\cdots\delta\varphi_{2}\left(t_{n_{\varphi_{1}}+1},\vx_{n_{\varphi_{1}}+1}\right)\cdots} = \\
&& \qquad\qquad\qquad 
\Gamma_{k}^{\left(n_{\varphi_{1}},n_{\varphi_{2}},\cdots\right)}\left(t_{1},\vx_{1},\cdots,t_{n_{\varphi_{1}}+1},\vx_{n_{\varphi_{1}}+1},\cdots\right)\,,
\end{eqnarray*}
where $n=\sum_i n_{\varphi_{i}}$ and $\varphi_i$ denote any of the fields contained in the multiplet $\varphi$. 
The Fourier transform of a vertex is denoted by
\begin{eqnarray*}
&& \Gamma_{k}^{\left(n_{\varphi_{1}},n_{\varphi_{2}},\cdots\right)}\left(\omega_{1},\vp_{1}\cdots\right) =  \\
&& \qquad\qquad\qquad
{\cal F}\left[\Gamma_{k}^{\left(n_{\varphi_{1}},n_{\varphi_{2}},\cdots\right)}\left(t_{1},\vx_{1}\cdots\right)\right]
\left(\omega_{1},\vp_{1}\cdots\right)\,
\end{eqnarray*}
with ${\cal F}$ the Fourier transform defined as Eq.~(\ref{eq:Fourier}) for each  variables $(\omega_i,\vp_i)$.
 A fully analogous  notation will be employed  for
the connected correlation functions $W_k^{(n)}$.

In order to take into account translation invariance in space and time, we also introduce the
 following generic $n$-point function:
\begin{eqnarray*}
&& \Gamma_{k}^{\left(n_{\varphi_{1}},n_{\varphi_{2}},\cdots\right)}\left(\omega_{1},\vp_{1}\cdots\right) 
= \left(2\pi\right)^{d+1}\delta\left(\sum_{i=1}^{n}\omega_{i}\right)\delta^d\left(\sum_{i=1}^{n}\vp_{i}\right) \\
&& \qquad\qquad\qquad
\bar{\Gamma}_{k}^{\left(n_{\varphi_{1}},n_{\varphi_{2}},\cdots\right)}\left(\omega_{1},\vp_{1}\cdots,\omega_{n-1},\vp_{n-1}\right)\,,
\end{eqnarray*}
where $\bar{\Gamma}_{k}^{\left(n\right)}$ hence denotes a 1-PI vertex ``stripped''
of the wavenumber and frequency  conserving delta functions. A fully analogous notation
is introduced for the $n$-point correlation function, $\bar{W}_{k}^{\left(n\right)}$.
In particular, a propagator stripped of the delta functions will
be written as $\bar{G}_{\varphi_{a}\varphi_{b}}\left(\omega,\vp\right)$.

In order not to overburden the notation in certain equations, we shall
sometimes employ the DeWitt notation. In this case, the average field
is denoted simply by $\varphi_{a}$ and the sources by $J^{a}$. The
summation convention is intended to 
mean $\varphi_{a}J^{a}\equiv\int_{t\vx}\varphi_{\alpha}\left(t,\vx\right)J^{\alpha}\left(t,\vx\right)$.

\subsection{Symmetries and Ward identities \label{subsec:Symmetries-and-WI}}

In this section we discuss the symmetries of the action (\ref{eq:S_final_total})
and the associated Ward identities for the EAA. It turns out
that it is particularly
useful to consider not only  the strict symmetries of the action but
also  the field transformations which leave the action invariant
up to terms linear in the fields since in this latter case one can
actually write down Ward identities which are even more constraining.
 We first  explicitly expound
 the case of the shift symmetry, and for the other symmetries, we only
 provide the field transformation, the action variation, and the ensuing Ward
identity. We refer the interested reader to Refs.~\onlinecite{Canet:2014cta,Tarpin:2018yvs} 
 where a similar analysis is detailed for the case of the NS field theory.
With abuse of notation, we denote the average fields by
employing the same symbols used for the fluctuating fields,
 i.e. $\varphi\equiv\left(\theta,\bar{\theta},v^{i},c,\bar{c}\right)$.

Before considering continuous symmetries, let us note that the action
(\ref{eq:S_final_total}) possesses the following $\mathbb{Z}_{2}$-symmetry:
$\theta\rightarrow-\theta$ and $\bar{\theta}\rightarrow-\bar{\theta}$.
An analogous symmetry is present in the ghost sector. Moreover, one can
 show that any term in the EAA must depend on at least one
 response field,  the EAA cannot have terms depending
on $\theta$ (or $c$) only, see Ref.~\onlinecite{Canet11b} for a detailed discussion
on this point.\\

\emph{Scalar field shifts. }Let us consider the transformation $\theta\left(t,\vx\right)\rightarrow\theta\left(t,\vx\right)+\varepsilon\left(t\right)$,
which is a time gauged shift of the advected scalar field. The variation
of the action reads
\begin{eqnarray*}
\delta S & = & \int_{t\vx}\,\bar{\theta}\left(t,\vx\right)\Biggr\{\partial_{t}+\frac{\kappa}{2}R_{k}\left(0\right)\Biggr\} \varepsilon\left(t\right)\,.
\end{eqnarray*}
By writing that this field transformation must leave the functional integral $Z$ unchanged, one obtains 
 the following Ward identity
\begin{eqnarray*}
&& \int_{t\vx}\,J_{\theta}\left(t,\vx\right)\varepsilon\left(t\right) = 
\Bigr\langle\int_{t\vx}\,\bar{\theta}\left(t,\vx\right)\Biggr\{\partial_{t}+\frac{\kappa}{2}R_{k}\left(0\right)\Biggr\} \varepsilon\left(t\right)\Bigr\rangle
\end{eqnarray*}
Expressing the  sources as functional derivatives of $\Gamma_k$ 
 using the definition (\ref{eq:Legendre}) of the Legendre transform then yields 
\begin{eqnarray*}
&& \int_{t\vx}\,\left(\frac{\delta\Gamma_{k}}{\delta\theta\left(t,\vx\right)}+\bar{\theta}\left(t,\vx\right)\frac{\kappa}{2}R_{k}\left(0\right)\right)\varepsilon\left(t\right) = \\
&&\quad\quad\quad\quad\quad\int_{t\vx}\,{\bar{\theta}\left(t,\vx\right)}\Biggr\{\partial_{t}+\frac{\kappa}{2}R_{k}\left(0\right)\Biggr\} \varepsilon\left(t\right) 
\end{eqnarray*}
Since this relation holds for an arbitrary infinitesimal $\varepsilon(t)$,
 this leads to the final Ward identity
\begin{equation}
 \int_{\vx}\,\frac{\delta\Gamma_{k}}{\delta\theta\left(t,\vx\right)} = -\int_{\vx}\,\partial_{t}\bar{\theta}\left(t,\vx\right)\,,\label{eq:WI-t-local-shift-1}
\end{equation}
which is local in time. The Ward identity (\ref{eq:WI-t-local-shift-1})
has an immediate interpretation: the term $\int\bar{\theta}\partial_{t}\theta$
in the microscopic action $S$ is not renormalized and it appears
  identically in the associated EAA. By taking further functional derivatives
of this identity with respect to the other fields, one obtains the following relations
in Fourier space
\begin{eqnarray}
&& \bar{\Gamma}^{(1,1,0,0,0)}_{k}(\omega_{1},\vp_1=0) =  i\omega_{1} \nonumber \\
&& \bar{\Gamma}_{k}^{\left(n_{\theta}\geq1,n_{\bar{\theta}},n_{v},n_{c},n_{\bar{c}}\right)}\left(\omega_{\theta},\vp_{\theta}=0,\cdots\right) = 0\,.\label{eq:WI-t-local-shift-2}
\end{eqnarray}
Equation (\ref{eq:WI-t-local-shift-2}) entails that any vertex having at least one vanishing wavenumber carried by a $\theta$
 field actually vanishes.\\

\emph{Response field shifts. }Let us consider the transformation $\bar{\theta}\left(t,\vx\right)\rightarrow\bar{\theta}\left(t,\vx\right)+\bar\varepsilon\left(t\right)$.
By exploiting the fact that the velocity field is divergenceless,
the variation of the action reads
\begin{eqnarray*}
\delta S & = & \int_{t\vx}\,\bar\varepsilon(t)\Biggr\{\partial_{t}\theta\left(t,\vx\right)+\frac{\kappa}{2} R_{k}\left(0\right)\theta\left(t,\vx\right)
-M_{k}\left(0\right)\bar{\theta}\left(t,\vx\right)\Biggr\}\,.
\end{eqnarray*}
One deduces that
\begin{eqnarray}
&& \Gamma_{k}^{\left(n_{\theta},n_{\bar{\theta}}\geq1,n_{v},n_{c},n_{\bar{c}}\right)}\left(\cdots,\omega_{\bar{\theta}},\vp_{\bar{\theta}}=0,\cdots\right) = 0\,.\label{eq:WI-t-local-Qb-shift-1}
\end{eqnarray}
which yields that any vertex having at least one vanishing wavenumber carried by a $\bar\theta$
 field also vanishes.\\

\emph{Time gauged Galilei transformation. }
We now consider an extended Galilean transformation, which corresponds to the infinitesimal field transformation
\begin{eqnarray}
v^{j}\left(t,\vx\right) & \rightarrow & v^{j}\left(t,\vx\right)-\varepsilon^{i}\left(t\right)\partial_{i}v^{j}\left(t,\vx\right)+\dot{\varepsilon}^{j}\left(t\right)\nonumber \\
\bar{\theta}\left(t,\vx\right) & \rightarrow & \bar{\theta}\left(t,\vx\right)-\varepsilon^{i}\left(t\right)\partial_{i}\bar{\theta}\left(t,\vx\right)\nonumber\\
\theta\left(t,\vx\right) & \rightarrow & \theta\left(t,\vx\right)-\varepsilon^{i}\left(t\right)\partial_{i}\theta\left(t,\vx\right)\,,\label{eq:Galilei-transformation}
\end{eqnarray}
and similarly for the ghost fields.
This transformation can be interpreted as a time-gauged extension of space translations.
The standard Galilean transformation is recovered for $\vec \varepsilon(t)= \vec \varepsilon \,t$.
 One finds
\begin{eqnarray*}
\delta S & = & \int_{t\vx}\,v^{i}\left(t,\vx\right)\left[\frac{1}{D_{0}}m^{d+\varepsilon}+\frac{1}{D_0} R_{k,vv}\left(0\right)\right]\p_t{\varepsilon}^{i}\left(t\right)\,.
\end{eqnarray*}
The associated Ward identity constrains the vertices for which one
of the velocity has a zero wavenumber. The explicit expression reads
\begin{eqnarray}
&& \Gamma_{\gamma_{1}\cdots\gamma_{n_{v}}\gamma_{n_{v}+1}}^{\left(n_{\theta},n_{\bar{\theta}},n_{v}+1,n_{c},n_{\bar{c}}\right)}\Bigr(\cdots,\underbrace{\omega_\ell,\vp_\ell=0}_{\ell={\rm velocity \;index}},\cdots\Bigr) = \nonumber \\
&& \qquad
-\sum_{i=1}^{n}\frac{p_{i}^{\gamma_\ell}}{\omega}\Gamma_{\gamma_{1}\cdots\gamma_{n_{v}} }^{\left(n_{\theta},n_{\bar{\theta}},n_{v},n_{c},n_{\bar{c}}\right)}\Bigr(\cdots,\underbrace{\omega_{i}+\omega,\vp_{i}}_{i^{{\rm th}}{\rm field}},\cdots\Bigr),\label{eq:WI-t-local-Galilei-1}
\end{eqnarray}
where $\gamma_{1}\cdots\gamma_{n_{v}+1}$ are the spatial
indices of the velocity fields, $\vp_{i}$ is the wavevector of the $i$th
field in the vertex, and $n=n_{\theta}+n_{\bar{\theta}}+n_{v}+n_{\theta}+n_{\bar{\theta}}$.\\

\emph{Time-gauged rotations}. The time-gauged version of spatial rotations
is also an extended symmetry of the action $S_{{\rm Kr}}$. This symmetry
was first identified for the two-dimensional NS equation, see  Ref.~\onlinecite{Tarpin:2018yvs}.
In the present work, we will not exploit this symmetry and leave its
study for future work.\\

\emph{BRS (Becchi-Rouet-Stora) symmetry}. We consider 
 the transformation $\theta\rightarrow\theta+\varepsilon c,\:\bar{c}\rightarrow\bar{c}-\varepsilon\bar{\theta}$
 where $\epsilon$ is now an anti-commuting Grassmann parameter.
 This transformation hence mixes the ``bosonic'' (scalar fields) and the ``fermionic'' (ghosts)
 sectors of the action. One observes that this transformation leaves the action invariant $\delta S=0$.
  It follows that
\begin{eqnarray*}
\int_{t\vx}\,\left[\frac{\delta\Gamma_{k}}{\delta\theta\left(t,\vx\right)}c\left(t,\vx\right)+\bar{\theta}\left(t,\vx\right)
\frac{\delta\Gamma_{k}}{\delta\bar{c}\left(t,\vx\right)}\right] & = & 0\,.
\end{eqnarray*}
By taking further functional derivatives one deduces relations between
 vertices involving ghosts. This is particularly useful to show that the kinetic
term of the velocity field is not renormalized\footnote{This non-renormalization of the velocity covariance
 can also shown using different arguments, as in the approach of
Ref.~\onlinecite{PhysRevE.58.1823}.} and to show that the terms in the EAA which depend on the velocity
alone are at most quadratic in the fields. 

Let us prove this last assertion, by explicitly
using the flow equation of the EAA. Let us assume that at the UV scale
 $k=\Lambda$, the EAA is such that $\Gamma_{\Lambda}\left[0,0,v,0,0\right]$
is at most quadratic in the velocity fields, which is the case  for the microscopic
action $S$ in (\ref{eq:S_final_total}).
Given this condition, we want to determine if the flow equation generates
terms which are of higher order in $v$. The associated flow equation
 reads
\begin{eqnarray*}
&& \partial_{s}\Gamma_{k}\left[0,0,v,0,0\right] = \frac{1}{2}\mbox{Tr}\left[\left(\Gamma_{k,vv}^{\left(2\right)}+R_{k}\right)^{-1}\dot{R}_{k,vv}\right]\\
&& \qquad
+\mbox{Tr}\left[\left(\Gamma_{k,\theta\bar{\theta}}^{\left(2\right)}+R_{k}\right)^{-1}\dot{R}_{k}\right]-\mbox{Tr}\left[\left(\Gamma_{k,c\bar{c}}^{\left(2\right)}+R_{k}\right)^{-1}\dot{R}_{k}\right]\\
&& \qquad
= \frac{1}{2}\mbox{Tr}\left[\left(\Gamma_{k,vv}^{\left(2\right)}+R_{k}\right)^{-1}\dot{R}_{k,vv}\right]\,,
\end{eqnarray*}
where the last equality ensues from the  BRS symmetry.
Since by assumption $\Gamma_{k,vv}^{\left(2\right)}$ has no dependence
on $v$, one finds that $\partial_{s}\Gamma_{k}\left[0,0,v,0,0\right]$ is
just a constant. Thus we can conclude that the quadratic-in-velocity nature of $\Gamma_{k}\left[0,0,v,0,0\right]$
 is preserved by the flow, i.e. it holds true at any scale $k$.

It is instructive to sketch how the non-renormalization of the kinetic
term of the velocity arises within the FRG approach. 
For this, one considers the flow equation for  $\Gamma^{(2)}_{k,v v}$,
 which can be obtained by taking two functional derivatives of the exact flow Eq. (\ref{eq:FRGE-generic_EAA})
  with respect to velocity fields.
  The vertices entering this flow equation are 3-point vertices with at least
   one velocity field (explicitly $\Gamma_k^{(1,1,1,0,0)}$, $\Gamma_k^{(0,2,1,0,0)}$, $\Gamma_k^{(0,0,1,1,1)}$ and 
   $\Gamma_k^{(0,0,3,0,0)}$), and 4-point vertices with at least two velocity fields
   (explicitly $\Gamma_k^{(1,1,2,0,0)}$, $\Gamma_k^{(0,2,2,0,0)}$, $\Gamma_k^{(0,0,2,1,1)}$ and 
   $\Gamma_k^{(0,0,4,0,0)}$). 
   The BRS Ward identities lead to the cancellation of all the diagrams with vertices 
    including the scalar fields and the ghosts. One is left  with only pure velocity vertices.
   However, since we have shown that $\Gamma_k$ is quadratic in $v$, it follows that
   $\Gamma_k^{(0,0,3,0,0)}=\Gamma_k^{(0,0,4,0,0)}=0$, and this implies the non-renormalization of
the velocity kinetic term.\\

\emph{Further symmetries in the ghost sector}. The ghost sector shares
the same symmetries as the scalar sector, i.e., shift and $\mathbb{Z}_{2}$
symmetries. Moreover the ghost action is also invariant under $\bar{c}\rightarrow e^{\alpha}\bar{c}$
and $c\rightarrow e^{-\alpha}c$.\\

\emph{Synthesis}. 
We have derived from the symmetries -- in an extended sense -- of the action,
 a set of Ward identities which will lead to the exact closure
 of the RG flow equation of any $n$-point correlation functions in the limit of large
 wavenumbers. The key feature that will be exploited is that
the identities (\ref{eq:WI-t-local-shift-2}),
(\ref{eq:WI-t-local-Qb-shift-1}), and (\ref{eq:WI-t-local-Galilei-1})
imply that any  1-PI $n$-point vertex carrying one vanishing wavevector
 is either zero or can be expressed in terms of lower-order $\left(n-1\right)$-point
vertices.

\section{Anomalous dimensions} \label{sec:preliminaries-and-non-renormalization-th}

In this section, we establish certain fundamental properties which are needed
 in the derivation of the closed flow equation for $n$-point correlation functions.

\subsection{Propagator and bare scaling dimensions}

The exact flow equation (\ref{eq:FRGE-generic_EAA})
involves as a fundamental ingredient the regularized propagator $\left(\Gamma_{k}^{\left(2\right)}+R_{k}\right)^{-1}$.
To obtain it, one writes the matrix $\big(\Gamma_{k}^{\left(2\right)} +R_{k}\big)$ of second functional
 derivatives of the EEA and regulator action. Since in Fourier space, it is diagonal in frequency and wavevector,
  one can then simply invert the matrix.
 Evaluated for vanishing values of the average field, $\varphi=0$, the regularized propagator
takes the following form in field space:
\begin{eqnarray}
\left(\bar{\Gamma}_{k}^{\left(2\right)}+R_{k}\right)^{-1} & = & \left(\begin{array}{ccccc}
\bar{G}_{\theta\theta} & \bar{G}_{\theta\bar{\theta}} & 0 & 0 & 0\\
\bar{G}_{\bar{\theta}\theta} & 0 & 0 & 0 & 0\\
0 & 0 & \bar{G}_{vv} & 0 & 0\\
0 & 0 & 0 & 0 & \bar{G}_{c\bar{c}}\\
0 & 0 & 0 & \bar{G}_{\bar{c}c} & 0
\end{array}\right)\,, \label{eq:propag}
\end{eqnarray}
with 
\begin{eqnarray}
\bar{G}_{\theta\bar\theta}(\omega,\vp) &=& \frac{1}{\Gamma_{k}^{\left(1,1,0,0,0\right)}(-\omega,\vp)+R_{k}(\vp)}\,, \nonumber\\
\bar{G}_{\theta\theta}(\omega,\vp) &=& -\frac{\Gamma_{k}^{\left(0,2,0,0,0\right)}(\omega,\vp)-M_{k}(\vp)}{\left|\Gamma_{k}^{\left(1,1,0,0,0\right)}(\omega,\vp)+R_{k}(\vp)\right|^{2}},
\end{eqnarray}
 and similarly for the ghost sector. $G_{vv}$ is the regularized velocity propagator, and
  since the velocity kinetic term is not renormalized,
 it writes
\begin{eqnarray}
\bar{G}_{v^{i}v^{j }}\left(\omega,\vp\right) & = & D_{0}\frac{P_{ij }(\vp)}{\left(p^{2}+m^{2}\right)^{\frac{d}{2}+\frac{\varepsilon}{2}}+R_{k,vv}\left(\vp\right)}\,.\label{eq:velocity-propagator}
\end{eqnarray}
Note that $m$ can be safely set to zero in this formalism thanks to the presence of the regulator.

Let us establish the bare scaling dimensions, which are defined
by the scaling properties of the fields as deduced from the microscopic
action (\ref{eq:S_final_total}). From the velocity kinetic term, it follows that the bare scaling
dimension of the velocity field should satisfy $TL^{d}L^{-d-\varepsilon}\left[v\right]^{2}=1$,
so that $\left[v\right]=L^{\varepsilon/2}/T^{1/2}$.\footnote{Note that this implies that the engineering dimension of the constant $D_{0}$
is not the same as its (trivial) scaling dimension.} For the quadratic term in $\theta$ and $\bar{\theta}$, one
 obtains that
$TL^{d}T^{-1}\left[\theta\bar{\theta}\right]$=1, which yields $\left[\theta\bar{\theta}\right]=L^{-d}$.
The dimension of the molecular viscosity reads $\left[\kappa\right]=L^{2}/T$.
The scaling dimension of $\bar{\theta}$ depends on the particular
form of the forcing covariance chosen. In the present case
we consider
\begin{eqnarray}
\int_{t\vx}\,\frac{1}{2}\bar{\theta} M_{k}\left(-\partial^{2}\right)\bar{\theta}
& \equiv & \int_{t\vx}\,\frac{1}{2}\bar{\theta} \left(-\partial^{2}\right)e^{\frac{\partial^{2}}{k^{2}}}\bar{\theta} \,,\label{eq:def-forcing-M_k}
\end{eqnarray}
to be dimensionless. The microscopic scaling dimension of $\bar{\theta}$
is then defined by $TL^{d}L^{-2}\left[\bar{\theta}\right]^{2}=1$
implying $\left[\bar{\theta}\right]=T^{-1/2}L^{\left(2-d\right)/2}$.

We parametrize the RG flow by introducing the following quantities:
the field renormalizations $\left(Z_{k,\theta},Z_{k,\bar{\theta}},Z_{k,v},Z_{k,c},Z_{k,\bar{c}}\right)$
and the running molecular diffusivity $\kappa_k/2$.
The precise definition of the field renormalization constants
in terms of the EAA is given in Sec.~\ref{subsec:Non-renormalization-theorem-and-anomalous-dim}.
For convenience, we also introduce a running coupling $\lambda_k$ for the three-point
vertex present in the bare action $\bar\theta v_i \partial_i \theta$, although 
  it is not renormalized (see below). Its bare scaling dimension is determined by
  $TL^{d}\left[\bar{\theta}\theta\right]\left[v\right]L^{-1}\left[\lambda_k\right]=1$,
   which leads to $[\lambda_k] = L^{1-\varepsilon/2}T^{-1/2}$.

\subsection{Non-renormalization theorems and anomalous dimensions \label{subsec:Non-renormalization-theorem-and-anomalous-dim}}

We associate to each field in the EEA the corresponding field renormalization $Z_{k,\varphi}^{1/2}$. 
Omitting the ghosts, the EAA then reads
\begin{eqnarray*}
&& \Gamma_{k} = 
\int_{t\vx}\,Z_{k,\bar{\theta}}^{1/2}\bar{\theta} \Biggr\{\partial_{t}+\lambda_k Z_{k,v}^{1/2}v^{i} \partial_{i}-\frac{\kappa_k}{2}\partial^{2}\Biggr\} Z_{k,\theta}^{1/2}\theta \\
& & \qquad\qquad\qquad
+\frac{1}{2}\int_{t\vx}\,Z_{k,v}v^{i} \frac{\left(-\partial^{2}+m^{2}\right)^{\frac{d}{2}+\frac{\varepsilon}{2}}}{D_{0}}v^{i} +\cdots\,,
\end{eqnarray*}
 The field renormalizations can thus be obtained through the relations
 \begin{align}
Z_{k,\bar{\theta}}^{1/2}Z_{k,\theta}^{1/2}&=\frac{\p}{\p\left(i\omega\right)}\Gamma_k^{\left(1,1,0,0,0\right)}\left(\omega=0,\vp=0\right)\nonumber\\
Z_{k,\bar{\theta}}^{1/2}Z_{k,\theta}^{1/2}\frac{\kappa_k}{2}&=\frac{\p^2}{\p p^2}\Gamma_k^{\left(1,1,0,0,0\right)}\left(\omega=0,\vp=0\right)\nonumber\\
\frac{Z_{k,v}}{D_0}&=\frac{\p}{\p p^{d+\varepsilon}}\Gamma_k^{\left(0,0,2,0,0\right)}\left(\omega=0,\vp=0\right)
\end{align}
The anomalous dimension of a field is then defined by $\eta_{\phi}\equiv-Z_{\phi}^{-1}\partial_{s}Z_{\phi}$,
 and the anomalous dimension associated to the molecular diffusivity
 by $\eta_{\kappa}\equiv-\kappa_k^{-1}\partial_{s}\kappa_k$.

 The Ward identities derived in
Sec.~\ref{subsec:Symmetries-and-WI} provide crucial constraints on these
  running constants. We now spell out the consequences
 of these identities and of the non-renormalization theorems presented in Sec.~\ref{subsec:Symmetries-and-WI}.
First,  we showed that the velocity kinetic term is not renormalized.
This immediately implies $Z_{k,v}=1$ and $\eta_{v}=0$. The Ward identity associated with the time-gauged
shifts Eq.~(\ref{eq:WI-t-local-shift-2})  
 implies $Z_{k,\bar{\theta}}^{1/2}Z_{k,\theta}^{1/2}=1$ and thus
$Z_{k,\theta}=Z_{k,\bar{\theta}}^{-1}$ and $\eta_{\theta}=-\eta_{\bar{\theta}}$.
 It follows that the Ward identity associated with the time-gauged Galilean symmetry Eq.~(\ref{eq:WI-t-local-Galilei-1})	imposes that this
   vertex is not renormalized,  and thus $\lambda_k=1$. 
Note that we redefine the cutoff action (\ref{eq:cutoff-action-Kraichnan})
such that the field renormalizations also appear in the cutoff
kernels. In the present case, this simply amounts to the subsitution
 $M_{k}\left(-\partial^{2}\right)\rightarrow Z_{k,\bar{\theta}}M_{k}\left(-\partial^{2}\right)$.

To study the existence of a fixed point, one usually
switches to a dimensionless formulation, by rescaling all quantities
  by suitable powers of the RG scale. In particular, dimensionless
wavevectors $\hat{\vp}$ are defined by $\hat{\vp}\equiv \vp/k$ and  dimensionless
frequencies  by $\hat{\omega}\equiv\omega/\left(\kappa_k k^{2}\right)$.
 At a fixed point, the RG flow of the dimensionless couplings vanishes.
In this respect, the non-renormalization of the coupling $\lambda_k$
has a striking consequence. Indeed, the associated dimensionless coupling
 reads $\hat{\lambda_k}\equiv\lambda_k k^{1-\varepsilon/2}\left(\kappa_k k^{2}\right)^{-1/2}=\lambda_k k^{-\varepsilon/2}\kappa_k^{-1/2}$.
Hence, since $\p_s \lambda_k=0$, its flow equation is simply given by the linear contribution
\begin{eqnarray}
\partial_{s}\hat{\lambda_k} & = & \left(-\frac{\varepsilon}{2}+\frac{\eta_{\kappa}}{2}\right)\hat{\lambda_k}\,.\label{eq:beta-lambda-dimless-kraichnan}
\end{eqnarray}
This implies that at a non-Gaussian fixed point, satisfying $\hat{\lambda}\neq0$, then $\eta_{\kappa}=\varepsilon$. 
 Given the relations $\eta_{\theta}=-\eta_{\bar{\theta}}$, $\eta_{\kappa}=\varepsilon$,
and $\eta_{v}=0$, one deduces that there is a single
anomalous dimension left to be determined at a fixed point.

\subsection{Energy budget}

In this section, we outline
an argument which allows one to fix the value of the remaining anomalous dimension $\eta_{\bar{\theta}}$.
In principle, in order to determine $\eta_{\bar{\theta}}$ within
the FRG framework, one has to derive the flow equations for the two-point functions $\Gamma_k^{(2)}$,
 integrate them, and read off the value of $\eta_{\bar{\theta}}$ at the fixed point.
 
 In this work, we shall employ a shortcut first proposed in  Ref.~\onlinecite{Canet:2014dta} for the case of NS equations.
Since we are interested in the stationary turbulent state,
 this requires that the mean energy rate injected by the forcing is constant and matches 
  the mean energy rate dissipated by the scalar field.
 The argument is thus based, in 3D,  on imposing a
finite value to $\langle f\theta\rangle$ in the ideal limit $L\rightarrow\infty$,
i.e. $k\rightarrow0$, or equivalently that $\langle\kappa\partial_{i}\theta\partial_{i}\theta\rangle$
 remains finite in the limit $\kappa\to 0$.

The averaged injected power  for the scalar can be estimated as
\begin{eqnarray}
&& \langle f\left(t,\vx\right)\theta\left(t,\vx\right)\rangle \propto
\int_{\vy}\Big\langle M_{k}\left(\vx-\vy\right)\bar{\theta}\left(t,\vy\right)\theta\left(t,\vx\right)\Big\rangle 
\label{eq:estimate-energy-inj-1} \\
&& = \int_{\omega \vq}M_{k}\left(\vq\right)G_{\theta\bar{\theta}}\left(\omega,\vq\right)\,\propto\,k^{d+2-\eta_{\bar{\theta}}}\int_{\hat{\omega}\hat{\vq}}\hat{M}_{k}\left(\hat{\vq}\right)\hat{G}_{\theta\bar{\theta}}\left(\hat{\omega},\hat{\vq}\right)\,. \nonumber
\end{eqnarray}
For $\langle f\theta\rangle$  to have a finite
limit when $k\to 0$ imposes $\eta_{\bar{\theta}}=d+2$. 
Let us remark that the estimate (\ref{eq:estimate-energy-inj-1})
is based on assuming standard scale invariance, in the sense that there is agreement
 between the RG scaling (in $k$) and the actual scaling properties of 1-PI vertices (in $\omega,\vq$).
However, although such correspondence is true in general in critical phenomena,
 it can be  violated in the case of turbulence, we refer to  Ref.~\onlinecite{Canet:2014dta}
for a discussion of this point in the context of the NS equations. Such
violations may be at the source of deviations from the Kolmogorov scaling.
 In this work,  we  limit ourselves to assuming standard
scale invariance to fix the anomalous dimension $\eta_{\bar{\theta}}$.
 This in turn yields the standard Obukhov-Corrsin scaling \cite{Obukhov49,Corrsin51}.

 Let us show that considering the energy rate dissipated by the 
  scalar leads to the same result.
 In the Kraichnan model, one may define the analog of the Kolmogorov
 dissipation scale by $\eta_{{\rm diss}}\sim\left(2\kappa/D_{0}^{\prime}\right)^{1/\varepsilon}$,
see eg.~Ref.~\onlinecite{Kupiainen:2006em}. The average dissipated power 
 can be estimated as
\begin{eqnarray*}
\kappa\langle\partial_{i}\theta\partial_{i}\theta\rangle & = & \kappa\int_{\omega}\int_{\vq, \,|\vq|<\eta_{{\rm diss}}^{-1}}q^{2}G_{\theta\theta}\left(\omega,\vq\right)\,,
\end{eqnarray*}
where $\eta_{{\rm diss}}^{-1}$ is a UV cutoff. The integral is dominated
by the UV contribution, which we can estimate by assuming standard
scale invariance. It follows that
\begin{eqnarray}
\lim_{\kappa\rightarrow0}\kappa\langle\partial_{i}\theta\partial_{i}\theta\rangle & \sim & \lim_{\kappa\rightarrow0}\kappa\left(\frac{1}{\eta_{{\rm diss}}}\right)^{d+2-\eta_{\bar{\theta}}+\eta_{\kappa}}\,.\label{eq:dissipation-limit-kappa_to_zero-and-kappa-power}
\end{eqnarray}
By inserting $\eta_{{\rm diss}}\sim\left(2\kappa/D_{0}^{\prime}\right)^{1/\varepsilon}$
and $\eta_{\kappa}=\varepsilon$, one deduces that reaching
 a finite limit requires $\eta_{\bar{\theta}}=d+2=5$.

\section{Correlation functions in the Kraichnan
model} \label{sec:closed-FRGE-Kraichnan}

The (connected) correlation functions $W_k^{(n)}$ can be expressed in terms
 of the 1-PI correlation functions $\Gamma_k^{(n)}$ via a sum of tree diagrams.
In the FRG framework, the $\Gamma_k^{(n)}$ are obtained by deriving
the associated flow equations, solving them, and eventually taking the limit
$k\rightarrow0$. For instance, by taking two functional derivatives
of the FRG flow equation (\ref{eq:FRGE-generic_EAA}), one obtains
an exact flow equation for the 2-point functions  $\Gamma_k^{(2)}$ (the inverse propagator).
 However, the RHS
of this flow equation depends on the $3$-point and $4$-point vertices,
which are governed by their own flow equation depending on higher-order vertices. 
It is then clear  that an infinite hierarchy is generated
and that a suitable approximation scheme must be implemented.

In the present section, we consider a closure scheme based on the Ward
identities detailed in Sec.~\ref{subsec:Symmetries-and-WI}. The
crucial approximation underlying this scheme is the following one. 
 All the vertices, which enter into the flow equation of a given vertex $\Gamma_k^{(n)}(\omega_i,\vp_i)$ ,
 are expanded in the loop wavevector $\vq\simeq 0$.
The rationale of this approximation is that the loop
wavevector is controlled by the cutoff derivative $\partial_{s}R_{k}\left(\vq\right)$,
 which vanishes  for $|\vq| \gtrsim k$.
 Hence, if one considers large external wave-vectors $\vp_i$, one may expand all the vertices
about zero $\vq$. This type of approximation is inspired by the 
Blaizot-M\'endez-Galain-Wschebor (BMW) scheme \cite{Blaizot:2005xy,Blaizot:2005wd,Blaizot:2006vr}.
It becomes exact in the limit where all $|\vp_i|\to \infty$.
In the context of  NS turbulence, this large wavenumber expansion was developed and analyzed
in  Ref.~\onlinecite{Canet:2014dta,Canet:2016nmg,Tarpin:2017uzn,Tarpin:2018yvs}.
The crucial feature of this scheme for the NS equations, which renders it particularly powerful,
 is that all the vertices with one wavector set to zero either vanish or are given exactly
  in terms of lower-order vertices through the Ward identities. 
  This results in a closure of the flow equation of any $\Gamma_k^{(n)}$, which 
   is thus expressed in terms of vertices with $m \leq n$ only, without
    any further approximation than the large wavenumber limit.
    We show in the following that this closure can also be achieved for the Kraichnan model,
     thanks to the Ward identities derived in Sec.~\ref{subsec:Symmetries-and-WI}.
 In order to do so, we closely follow the derivation detailed in Ref.~\onlinecite{Tarpin:2017uzn}
 to apply it to the Kraichnan model. Since the calculations which lead to
the final form of the closed flow equation are formally identical, we limit
ourselves to outlining the main steps of the derivation and refer to  Ref.~\onlinecite{Tarpin:2017uzn}
for details. In this work, we derive the general closed flow equation
 for a generic $n$-point correlation function, but we only focus on the solution of 
 the 2-point function of the scalar field.

\subsection{Closed flow equations for $n$-point correlation functions\label{subsec:Closed-flow-equations}}

For this derivation, it is more convenient to consider the exact FRG equation for
 the generating functional of the connected correlation functions
 $W_{k}\left[J\right]$, which  reads \cite{Delamotte:2007pf}
\begin{eqnarray*}
\partial_{s}W_{k}\left[J\right] & = & -\frac{1}{2}\partial_{s}R_{k,ij}\left[\frac{\delta^{2}W_{k}\left[J\right]}{\delta J^{i}\delta J^{j}}+\frac{\delta W_{k}\left[J\right]}{\delta J^{i}}\frac{\delta W_{k}\left[J\right]}{\delta J^{j}}\right]\,,
\end{eqnarray*}
where we use deWitt notation, in particular the integrals are implicit. By taking  functional derivatives
 of this equation,
one obtains the exact flow equations for $n$-point correlation functions as
\begin{eqnarray}
&& \partial_{s}\frac{\delta^{n}W_{k}\left[J\right]}{\delta J^{1}\cdots\delta J^{n}} = 
-\frac{1}{2}\partial_{s}R_{k,ij}\Biggr[\frac{\delta^{n+2}W_{k}\left[J\right]}{\delta J^{i}\delta J^{j}\delta J^{1}\cdots\delta J^{n}} \nonumber \\
&& \quad
+\sum_{\stackrel{\left(\left\{ a_{k}\right\} ,\left\{ a_{\ell}\right\} \right)}{k+\ell=n}}\frac{\delta^{k+1}W_{k}\left[J\right]}{\delta J^{i}\delta J^{a_{1}}\cdots\delta J^{a_{k}}}\frac{\delta^{\ell+1}W_{k}\left[J\right]}{\delta J^{j}\delta J^{a_{k+1}}\cdots\delta J^{a_{k+\ell}}}\Biggr]\,,\label{eq:FRGE-W(n)-1}
\end{eqnarray}
where $\left(\left\{ a_{k}\right\} ,\left\{ a_{\ell}\right\} \right)$
indicates all possible bipartions of the indices $1,\cdots,n$.

Let us first consider the second term in Eq.~(\ref{eq:FRGE-W(n)-1}).
It is straightforward to check that in wavevector space the cutoff derivative
term $\partial_{s}R_{k,ij}$ is a function of the total wavevector
carried by $J^{a_{1}},\cdots,J^{a_{k}}$, say $\vp\equiv \vp_{a_{1}}+\cdots+ \vp_{a_{k}}$.
In the large wavenumber limit, the cutoff derivative $\partial_{s}R_{k,ij}\left(\vp\right)$
is exponentially suppressed. Therefore, in this limit
 the second term of Eq.~(\ref{eq:FRGE-W(n)-1}) is negligible and one can discard it.

Let us now turn to the first term in Eq.~(\ref{eq:FRGE-W(n)-1}).
We first rewrite it in terms of functional derivatives with respect
to the average field $\varphi$, rather than functional derivatives
with respect to the sources, since the former will generate 1-PI vertices to which
one can apply the intended approximation scheme ($\vq$-expansion and Ward identities). 
This leads to
\begin{eqnarray}
&& \partial_{s}R_{k,ij}\frac{\delta^{n+2}W_{k}\left[J\right]}{\delta J^{i}\delta J^{j}\delta J^{1}\cdots\delta J^{n}} = \nonumber \\
&& \qquad\quad
-\frac{1}{2}\partial_{s}R_{k,ab}\frac{\delta\varphi_{i}}{\delta J^{a}}\frac{\delta\varphi_{j}}{\delta J^{b}}\frac{\delta}{\delta\varphi_{i}}\frac{\delta}{\delta\varphi_{j}}\frac{\delta^{n}W_{k}\left[J\right]}{\delta J^{1}\cdots\delta J^{n}}
\nonumber \\
&& \qquad\quad 
-\frac{1}{2}\partial_{s}R_{k,ab}\left(\frac{\delta^{3}W_{k}\left[J\right]}{\delta J^{a}\delta J^{b}\delta J^{j}}\right)\frac{\delta}{\delta\varphi_{j}}\frac{\delta^{n}W_{k}\left[J\right]}{\delta J^{1}\cdots\delta J^{n}}\,.\label{eq:FRGE-W(n)-step-2}
\end{eqnarray}
At vanishing sources and for corresponding vanishing average fields,
the  term $\p_s R_k W_k^{(3)}$ in (\ref{eq:FRGE-W(n)-step-2}) is proportional to
the flow of $\frac{\delta W_{k}}{\delta J^{j}}=\varphi_{j}$, which
itself vanishes. Thus, one concludes that also the second term in
(\ref{eq:FRGE-W(n)-step-2}) vanishes. The flow equation
hence takes the form
\begin{eqnarray}
\partial_{s}\frac{\delta^{n}W_{k}\left[J\right]}{\delta J^{1}\cdots\delta J^{n}} & = & 
-\frac{1}{2}\partial_{s}R_{k,ab}\frac{\delta\varphi_{i}}{\delta J^{a}}\frac{\delta\varphi_{j}}{\delta J^{b}}\frac{\delta^2}{\delta\varphi_{i}\delta\varphi_{j}}\frac{\delta^{n}W_{k}\left[J\right]}{\delta J^{1}\cdots\delta J^{n}}\,,
\qquad
\label{eq:FRGE-W(n)-3}
\end{eqnarray}
where the RHS must be though of as a functional of the average field
which  is eventually evaluated at $\varphi=0$.

At this point, one can proceed and evaluate the functional derivatives
$\frac{\delta}{\delta\varphi_{i}}$ and $\frac{\delta}{\delta\varphi_{j}}$
in (\ref{eq:FRGE-W(n)-3}). In order to implement the closure strategy,
we observe that each of these functional derivative carries a wavevector $\vq$
which is controlled by the cutoff derivative $\partial_{s}R_{k,ab}\left(\vq\right)$.
In turn this implies that  $\frac{\delta}{\delta\varphi_{i}}$ will generate
a 1-PI vertex which has its wavevector carried by $\varphi_i$ equal to $\vq$.
The  large wavenumber expansion then leads to setting  $\vq=0$ in
 all such vertices. One can then apply the Ward identities studied in 
Sec.~\ref{subsec:Symmetries-and-WI}. They
have an immediate stricking consequence: the only functional derivatives
$\frac{\delta}{\delta\varphi_{i}}$ which lead to a non-zero contribution
are those with respect to the velocity field only, since when $\vq$ is associated
to any other fields, then the Ward identities yield that the corresponding
vertex vanishes.
Therefore, one can further simplify the flow equation, which takes
the form
\begin{eqnarray}
\partial_{s} W^{\left(n\right)}_{k} 
& = & 
-\frac{1}{2}\left(G_{v^{i}v^{\alpha}}\partial_{s}R_{k,v^{\alpha}v^{\beta}}G_{v^{\beta}v^{j}}\right)\frac{\delta^2}{\delta v^{i} \delta v^{j}}
W^{\left(n\right)}_{k}  \,. \qquad
\label{eq:FRGE-W(n)-4}
\end{eqnarray}
As it is, Eq.~(\ref{eq:FRGE-W(n)-4}) is not closed yet. More
explicitly, this equation reads
\begin{eqnarray*}
\partial_{s} W^{\left(n\right)}_{k}  & = & -\frac{1}{2}\int_{\omega \vq}\left(\bar{G}_{v^{i}v^{\alpha}}\partial_{s}R_{k,v^{\alpha}v^{\beta}}\bar{G}_{v^{\beta}v^{j}}\right)\left(\omega,\vq\right) \\
&& \times
\frac{\delta}{\delta v^{i}\left(-\omega,-\vq\right)}\frac{\delta}{\delta v^{j}\left(\omega,\vq\right)}W^{\left(n\right)}_{k} \,.
\end{eqnarray*}
However,  acting on $W_{k}^{\left(n\right)}$ with $\frac{\delta}{\delta v}$
generates 1-PI vertices which are controlled by (\ref{eq:WI-t-local-Galilei-1})
after setting $\vq=0$. 
 One can show that the action of the velocity functional derivative yields
\begin{eqnarray}
&& \left(\frac{\delta}{\delta v^{i}\left(\omega,\vq\right)}W^{\left(n\right)}_{k} \left(\omega_{1},\vp_{1},\cdots \right)\right)\Bigr|_{J=0, \vq=0} \nonumber \\
&& = \sum_{k=1}^{n}-\frac{p_{k}^{i}}{\omega}W_{k}^{\left(n\right)}\left(\omega_{1},\vp_{1},\cdots,\omega_{k}+\omega,\vp_{k},\cdots\right)\nonumber \\
&& \equiv {\cal D}^{i}\left(\omega\right) W_{k}^{\left(n\right)}\left(\omega_{1},\vp_{1},\cdots\right) \,.\label{eq:one-q-zero-v-derivatives}
\end{eqnarray}
 and  a further functional derivative can be expressed as
\begin{eqnarray}
&& \frac{\delta}{\delta v^{i}\left(-\omega,-\vq\right)}\frac{\delta}{\delta v^{j}\left(\omega,\vq\right)}W_{k}^{\left(n\right)}\left(\omega_{1},\vp_{1},\cdots\right)\Bigr|_{\vq=0} \nonumber \\
&& \qquad\qquad\quad = 
{\cal D}^{i}\left(-\omega\right){\cal D}^{j}\left(\omega\right)W_{k}^{\left(n\right)}\left(\omega_{1},\vp_{1},\cdots\right)\,.\label{eq:two-q-zero-v-derivatives}
\end{eqnarray}
Note that $\vq$ is set to zero only at the level of vertices (not in the $W_k^{(n)}$)
 and also the Ward identities are applied to these vertices. 
 The proof of the resulting identities (\ref{eq:one-q-zero-v-derivatives}) and (\ref{eq:two-q-zero-v-derivatives})
  at the level of the  correlation functions $W_k^{(n)}$
  is quite lengthy, but it poses no difficulty 
 as it is strictly similar to  Ref.~\onlinecite{Tarpin:2017uzn}, to which
we refer the reader for the detailed demonstration.

Using the identities (\ref{eq:one-q-zero-v-derivatives}) and (\ref{eq:two-q-zero-v-derivatives}) to express 
  the RHS of Eq.~(\ref{eq:two-q-zero-v-derivatives})
then leads to the final expression for the flow equation of $W_k^{n}$ at large wavenumbers
\begin{eqnarray}
&& \partial_{s}W_{k}^{n}\left(\cdots,\omega_{k},\vp_{k},\cdots\right) 
=
\frac{1}{2}\int_{\omega \vq}\left(\bar{G}_{v^{i}v^{\alpha}}\partial_{s}R_{k,v^{\alpha}v^{\beta}}\bar{G}_{v^{\beta}v^{j}}\right)\left(\omega,\vq\right)
\qquad
\nonumber \\
&& \quad
\sum_{k,\ell=1}^{n}\frac{p_{k}^{i}p_{\ell}^{j}}{\omega^{2}}W_{k}^{\left(n\right)}\left(\cdots,\omega_{k}+\omega,\vp_{k},\cdots,\omega_{\ell}-\omega,\vp_{\ell},\cdots\right)\,,\label{eq:FRGE-W(n)-5}
\end{eqnarray}
which does not involved any higer-order correlation functions $W_{k}^{\left(m>n\right)}$,
showing that it is closed.

\subsection{Two-point correlation function in the large wavenumber limit \label{subsec:2pt-function-in-large-P}}

We now specify to the  flow equation for $\bar{G}_{\theta\theta}\left(\omega,\vp\right)=\bar{W}_{k}^{\left(2,0,0,0,0\right)}\left(\omega,\vp\right)$.
Starting from equation (\ref{eq:FRGE-W(n)-5}), a straightforward
calculation leads to
\begin{eqnarray*}
&& \partial_{s}\bar{G}_{\theta\theta}\left(\omega_{1},\vp\right) = \frac{1}{2}\int_{\omega \vq}\left(\bar{G}_{v^{i}v^{\alpha}}\partial_{s}R_{k,v^{\alpha}v^{\beta}}\bar{G}_{v^{\beta}v^{j}}\right)\left(\omega, \vq\right) \times \\
&&
\frac{p^{i}p^{j}}{\omega^{2}}\Biggr\{2\bar{G}_{\theta\theta}\left(\omega_{1}, \vp\right)-\bar{G}_{\theta\theta}\left(\omega_{1}+\omega,\vp\right)-\bar{G}_{\theta\theta}\left(\omega_{1}-\omega,\vp\right)\Biggr\}\,.
\end{eqnarray*}
By inverse Fourier transforming  to real time, one obtains
\begin{eqnarray}
&& \partial_{s}\bar{G}_{\theta\theta}\left(t,\vp\right) = \frac{1}{2}\int_{\omega \vq}\left(\bar{G}_{v^{i}v^{\alpha}}\partial_{s}R_{k,v^{\alpha}v^{\beta}}\bar{G}_{v^{\beta}v^{j}}\right)\left(\omega,\vq\right)p^{i}p^{j} \nonumber \\
&& \qquad\qquad\qquad\qquad\quad
\times \Biggr\{\frac{2-2\cos\left(t\omega\right)}{\omega^{2}}\Biggr\}\bar{G}_{\theta\theta}\left(t,\vp\right)\,.\label{eq:FRGE-Gqq(tp)-1}
\end{eqnarray}
Equation (\ref{eq:FRGE-Gqq(tp)-1}) is written in terms of the scalar
field and the velocity propagators. At this point, one of the peculiarities
of the Kraichnan model intervenes, namely that the velocity propagator
is known exactly, Eq.~(\ref{eq:velocity-propagator}), and
it does not depend on the frequency. This allows one to perform
 explicitly the integration over the internal frequency, which gives
\begin{eqnarray}
&&
\partial_{s}\bar{G}_{\theta\theta}\left(t,\vp\right) = \frac{d-1}{2d}\int_{\vq}\frac{D_{0}}{\left(\left(q^{2}+m^{2}\right)^{\frac{d}{2}+\frac{\varepsilon}{2}}+R_{k,vv}\right)^{2}}\partial_{s}R_{k,vv}\left(\vq\right) \nonumber \\
&& \qquad\qquad\qquad\qquad\qquad\qquad\qquad\quad
\times p^{2}\left|t\right|\bar{G}_{\theta\theta}\left(t,\vp\right)\,.
\label{eq:kraichnan-dimful-FRGE}
\end{eqnarray}

We are interested in the fixed-point solution of this equation.
It  is thus convenient to switch
to dimensionless variables. We thus introduce   $\hat{\vp}=\vp/k$,
$\hat{t}\equiv t\kappa_k k^{2}$, 
\begin{eqnarray*}
\bar{G}_{\theta\theta}\left(t,\vp\right) & = & k^{\eta_{\kappa}-\eta_{\bar{\theta}}}\hat{\bar{G}}_{\theta\theta}\left(\hat{t},\hat{\vp}\right)\,,
\end{eqnarray*}
and the dimensionless regulator following $R_{k,vv}(\vq) = k^{d+\epsilon}\hat{R}_{k,vv}(\hat{\vq})$, and thus 
$$ \p_s R_{k,vv}(\vq) = k^{d+\epsilon}\Big((d+\epsilon)\hat{R}_{k,vv}(\hat{\vq})-\hat{q}\p_{\hat{q}}\hat{R}_{k,vv}(\hat{\vq})\Big).$$
We define
\begin{eqnarray*}
\hat {\alpha}_k & \equiv & \frac{d-1}{2d}
\int_{\hat{\vq}}\frac{\partial_{s}\hat{R}_{k,vv}\left(\hat{\vq}\right)}{\left(\left(\hat{q}^{2}+\hat{m}^{2}\right)^{\frac{d}{2}+\frac{\varepsilon}{2}}+\hat{R}_{k,vv}\right)^{2}}\,,
\end{eqnarray*}
which, at the fixed point, becomes a pure number that can be calculated given any explicit cutoff kernel $R_{k,vv}$, and eg. $m=0$.
 The dimensionless flow equation then reads
\begin{eqnarray}
&& \left(\partial_{s}+\left(\eta_{\kappa}-\eta_{\bar{\theta}}\right)+\left(2-\eta_{\kappa}\right)\hat{t}\partial_{\hat{t}}-\hat{\vp}\partial_{\hat{\vp}}\right)\hat{\bar{G}}_{\theta\theta}\left(\hat{t},\hat{\vp}\right) 
=  \nonumber\\
&&\qquad\qquad\qquad\qquad\qquad 
= D_{0}  \hat {\alpha}_k \hat{\lambda_k}^{2}\hat{p}^{2}\left|\hat{t}\right|\hat{\bar{G}}_{\theta\theta}\left(\hat{t},\hat{\vp}\right)\,\nonumber\\
\label{eq:dimless-FRGE-Gqq-1}
\end{eqnarray}
where $\hat{\lambda}_k$ is the dimensionless coupling associated to the non-renormalized bare one
$\lambda_k=1$. 
By inserting $\eta_{\kappa}=\varepsilon$ and $\eta_{\bar{\theta}}=5$, and considering the fixed-point
 (i.e. $\p_s=0$ in Eq.~(\ref{eq:dimless-FRGE-Gqq-1})), one obtains
\begin{eqnarray}
&& \left(-\left(5-\varepsilon\right)+\left(2-\varepsilon\right)\hat{t}\partial_{\hat{t}}-\hat{\vp}\partial_{\hat{\vp}}\right)\hat{\bar{G}}_{\theta\theta}\left(\hat{t},\hat{\vp}\right)\nonumber \\
&& \qquad\qquad\qquad\qquad\qquad 
= D_{0}\hat{\alpha}\hat{\lambda}^{2}\hat{p}^{2}\left|\hat{t}\right|\hat{\bar{G}}_{\theta\theta}\left(\hat{t},\hat{\vp}\right)\,.\label{eq:FP-FRGE-Gqq-1}
\end{eqnarray}
where $\hat{\lambda}$ and $\hat{\alpha}$ are the fixed-point values of $\hat{\lambda}_k$ and $\hat\alpha_k$.

At equal time $\hat{t}=0$, Eq.~(\ref{eq:FP-FRGE-Gqq-1})
takes a very simple form since the RHS and the logarithmic
$\hat{t}$-derivative vanish, and the solution is a power law 
\begin{eqnarray*}
\hat{\bar{G}}_{\theta\theta}\left(0,\hat{\vp}\right) & \sim & \hat{p}^{-\left(5-\varepsilon\right)}\,.
\end{eqnarray*}
This power law  corresponds to the scaling expected in the inertial
range ${\bar{G}}_{\theta\theta}\left(0,{x}\right)-{\bar{G}}_{\theta\theta}\left(0,0\right)\sim x^{2-\varepsilon}$.

The general solution of Eq.~(\ref{eq:FP-FRGE-Gqq-1}) is given
by
\begin{eqnarray}
\hat{\bar{G}}_{\theta\theta}\left(\hat{t},\hat{\vp}\right) & = & \hat{p}^{-\left(5-\varepsilon\right)}e^{-D_{0}\frac{\hat{\alpha}}{\varepsilon}\hat{\lambda}^{2}\hat{p}^{2}|\hat{t}|}\hat{f}\left(\hat{t}\hat{p}^{2-\varepsilon}\right)\,,\label{eq:dimless-FP-sol-Gqq-1}
\end{eqnarray}
where $f\left(\hat{t}\hat{p}^{2-\varepsilon}\right)$ is a universal scaling
function, which can be determined from FRG methods. To do so, 
 it is not sufficient to solve the fixed-point equation but one has
 to evolve the flow equation from an initial condition corresponding to the bare action (\ref{eq:S_final_total}) down
to $k=0$ (in fact one can stop at the scale $k$ where the fixed point is reached). 
The exponential present in the solution (\ref{eq:dimless-FP-sol-Gqq-1})
does not have a smooth limit as $\varepsilon\rightarrow0$. This is
not surprising since  the velocity two-point correlation function
(\ref{eq:vel-2pt-correlation-function}) itself is not well defined
as $\varepsilon\rightarrow0$. In terms of the parameter $D_{0}^{\prime}$
the solution has a smooth behavior in $\varepsilon$ (note  that
$\hat{\alpha}$ is also a function of $\varepsilon$) which reads,
switching back to the dimensionful physical quantities,
\begin{eqnarray}
\bar{G}_{\theta\theta}\left(t,\vp\right)
& = & p^{-\left(5-\varepsilon\right)}e^{-D_{0}^{\prime}\hat{\alpha} k^{-\varepsilon}p^{2}|t|} {f}\left(tp^{2-\varepsilon}\right)\,.\label{eq:dimless-FP-sol-Gqq-3}
\end{eqnarray}
Note that the exponential in (\ref{eq:dimless-FP-sol-Gqq-3}) depends
explicitly on $k$. This is a sign of the breaking of standard
scale invariance. When considering the fixed point  for $k\rightarrow0$,
one typically identifies the RG scale $k$ with the inverse of the energy injection
scale, $k=L^{-1}$ since the flow essentially stops when crossing this scale.
However, for the Kraichnan model, it makes sense
to identify $k$ with the mass scale present in the velocity propagators
since the coefficient $\hat{\alpha}$ is fully determined by it. 
This yields
\begin{eqnarray}
\bar{G}_{\theta\theta}\left(t,\vp\right) & = & p^{-\left(d+2-\varepsilon\right)}e^{-D_{0}^{\prime}\hat{\alpha} m^{-\varepsilon}p^{2}|t|}f\left(tp^{2-\varepsilon}\right)\,.\label{eq:dimless-FP-sol-Gqq-4}
\end{eqnarray}
which is one of the main results of this work. Let us emphasize that this expression
 is valid at large wavenumber $p$ but for arbitrary time delays $t$ in the stationary state. The temporal decorrelation of the scalar
  in the Kraichan model is always exponential at any time scale. This is due to the complete decorrelation in time of the stochastic velocity field in this model. One can  indeed remark that if this feature is relaxed by introducing some time correlation in the covariance Eq.~(\ref{eq:vel-2pt-correlation-function}), 
   the expression (\ref{eq:dimless-FP-sol-Gqq-4}) is no longer valid at small $t$, but is replaced by a Gaussian decay  as occurs for the real scalar, see Sec.~\ref{sec:closed-FRGE-for-advected-scalars}.

\subsection{Two-point correlation function from the Dyson equation \label{subsec:2pt-correlation-function-from-Dyson-eq}}

The simplifying assumptions on the velocity field in the Kraichnan model
 lead to the important consequence that the equal-time
two-point correlation function $\langle\theta\left(t,x\right)\theta\left(t,y\right)\rangle$
can be computed exactly, see Ref.~\onlinecite{RevModPhys.73.913} and references
therein. Such an exact solution is obtained by deriving a closed form of
the associated Hopf equation, which allows one to write  an exact
differential equation for the two-point correlation function,
 see, eg. Refs.~\onlinecite{RevModPhys.73.913,Falkovich1996,Eyink2013}.

The connection between this exact differential equation for the two-point
 correlation function and the field theoretical calculation was worked out in
Ref.~\onlinecite{PhysRevE.58.7381}, which showed how to retrieve
the differential equation from the Dyson equation
 in a steady state. 
In this section,
we follow the approach used in  Ref.~\onlinecite{PhysRevE.58.7381} and extend it
to compute the two-point Eulerian correlation function of the scalar \emph{at unequal times},
 thereby allowing for a comparison
 with the result obtained by FRG methods, Eq.~(\ref{eq:dimless-FP-sol-Gqq-4}).

The aim is to compute the self-energies, which can be defined from the propagators as
\begin{eqnarray*}
\bar{G}_{\theta \bar{\theta}}\left(\varpi,\vp\right) & \equiv & \frac{1}{-i\varpi+\frac{\kappa}{2}p^{2}-\Sigma_{\theta\bar{\theta}}\left(\varpi,\vp\right)}\\
\bar{G}_{\theta\theta}\left(\varpi,\vp\right) & \equiv & \frac{\left(M_{k}\left(\vp\right)+\Sigma_{\bar{\theta}\bar{\theta}}\left(\varpi,\vp\right)\right)}{\left|-i\varpi+\frac{\kappa}{2}p^{2}-\Sigma_{\theta\bar{\theta}}\left(\varpi,\vp\right)\right|^{2}}\,.
\end{eqnarray*}
We consider the one-loop diagrams
 which enter in the determination of $\Sigma_{\bar{\theta}\theta}$
and $\Sigma_{\theta\theta}$, and express them in terms of the exact
propagators. Using the expression (\ref{eq:vel-2pt-correlation-function}) for the velocity propagator,
one finds \cite{PhysRevE.58.7381}
\begin{eqnarray*}
\Sigma_{\theta\bar{\theta}}\left(\varpi,\vp\right) & = & 
-\int_{\omega \vq}\bar{G}_{\theta\bar{\theta}}\left(\omega+\varpi,\vq+\vp\right)
p^{\alpha}\bar{G}_{v^{\alpha}v^{\beta}}\left(\vq\right)(p^{\beta}+q^{\beta})\\
 & = & -p^{2}\left[D_{0}\frac{\left(d-1\right)}{2d}\frac{\Gamma\left(\varepsilon/2\right)m^{-\varepsilon}}{\left(4\pi\right)^{d/2}\Gamma\left(d/2+\varepsilon/2\right)}\right]\,
\end{eqnarray*}
This a striking feature of the Kraichnan model: 
The expression of $\Sigma_{\bar{\theta}\theta}\left(\varpi,\vp\right)$
 is \emph{frequency independent} because the velocity propagator is
itself frequency independent, allowing one to  shift the frequency
in the integral. The frequency integration  simply yields
$\int_{\omega}\bar{G}_{\theta\bar{\theta}}(\omega,\vp)=\frac{1}{2}$,
 and the wavevector integral can then also be carried out, leading
  to  an extremely simple $\vp$-dependence as well.
The 1-PI vertex $\Gamma^{\left(1,1,0,0,0\right)}$ then takes the following
simple form
\begin{eqnarray*}
&&
\Gamma^{\left(1,1,0,0,0\right)}\left(\varpi,\vp\right) \\
 &&= i\varpi + p^2\left[\frac{\kappa}{2} + D_{0}\frac{\left(d-1\right)}{2d}\frac{\Gamma\left(\varepsilon/2\right)m^{-\varepsilon}}{\left(4\pi\right)^{d/2}\Gamma\left(d/2+\varepsilon/2\right)}\right]\\
 && \equiv i\varpi  + p^2\frac{\kappa_{{\rm ren}}}{2}\, ,
\end{eqnarray*}
where we introduced the renormalized molecular viscosity $\kappa_{{\rm ren}}$. 

The self-energy of the response field is given by
\begin{eqnarray*}
&& 
\Sigma_{\bar{\theta}\bar{\theta}}\left(\varpi,\vp\right) \\
&& 
= \int_{\omega \vq}\bar{G}_{v^{\alpha}v^{\beta}}\left(\vq\right)
\left(p^{\alpha}+q^{\alpha}\right)(p^{\beta}+q^{\beta})G_{\theta\theta}\left(\omega+\varpi,\vq+\vp\right)\\
&& 
= \int_{\omega \vq}\frac{D_{0}}{\left(q^{2}+m^{2}\right)^{\frac{d+\varepsilon}{2}}}\left(p^{2}-\frac{\left(\vp\cdot \vq\right)^{2}}{q^{2}}\right) \\
&& \qquad\qquad\qquad\qquad
\times \frac{M_{k}\left(\vq+\vp\right)+\Sigma_{\bar{\theta}\bar{\theta}}\left(\varpi+\omega,\vq+\vp\right)}{\left(\omega+\varpi\right)^{2}
+\left(\frac{\kappa_{{\rm ren}}}{2}\left(\vq+\vp\right)^{2}\right)^{2}}\,.
\end{eqnarray*}
 A shift in the integration frequency shows
that this self-energy is also independent of $\varpi$. It is clear that
the frequency independence of the self-energies is a peculiarity of
the Kraichnan model.

At this point, one can express the two-point correlation function as
\begin{eqnarray*}
\bar{G}_{\theta\theta}\left(\varpi,\vp\right) &=& 
\frac{1}{\varpi^{2}+\left(\frac{\kappa_{{\rm ren}}}{2}p^{2}\right)^{2}}
\Biggr(M_{k}\left(\vp\right)+\int_{\vq}\frac{D_{0}}{\left(q^{2}+m^{2}\right)^{\frac{d+\varepsilon}{2}}} \\
& \times &
\left(p^{2}-\frac{\left(\vp\cdot \vq\right)^{2}}{q^{2}}\right)\frac{M_{k}\left(\vq+\vp\right)
+\Sigma_{\bar{\theta}\bar{\theta}}\left(\vq+\vp\right)}{\kappa_{{\rm ren}}\left(\vq+\vp\right)^{2}} \Biggr)\\
& \equiv & \frac{F\left(\vp\right)}{\varpi^{2}+\left(\frac{\kappa_{{\rm ren}}}{2}p^{2}\right)^{2}}\,,
\end{eqnarray*}
where $F\left(\vp\right)$ is simply a shorthand for the expression in the
parentheses. It is straightforward to transform back to real time and
obtain
\begin{eqnarray}
\bar{G}_{\theta\theta}\left(t,\vp\right) & = & e^{-\frac{\kappa_{{\rm ren}}}{2}p^{2}\left|t\right|}\frac{F\left(\vp\right)}{\kappa_{{\rm ren}}p^{2}}\,,\label{eq:Gqq-t-p-via-Dyson-fin}
\end{eqnarray}
which provides another expression for the two-point correlation function of the scalar, that we compare
 with the FRG one in the following section.

\subsection{Comparison of the two approaches and remarks}

Let us comment on the two expressions  we have derived for $\bar{G}_{\theta\theta}\left(t,\vp\right)$.
First of all, the temporal dependence
of (\ref{eq:Gqq-t-p-via-Dyson-fin}) is of the same form as the one obtained
by FRG methods, namely an exponential with an exponent proportional
to $p^{2}\left|t\right|$. It is also interesting to compare the coefficients
appearing in front of this factor. In (\ref{eq:Gqq-t-p-via-Dyson-fin})
the coefficient $\frac{\kappa_{{\rm ren}}}{2}$ has two contributions:
one is just the microscopic molecular diffusivity and the other reads
\begin{eqnarray}
\frac{\kappa_{{\rm ren}}}{2}-\frac{\kappa}{2} & = & D_{0}\frac{\left(d-1\right)}{2d}\frac{\Gamma\left(\varepsilon/2\right)m^{-\varepsilon}}{\left(4\pi\right)^{d/2}\Gamma\left(d/2+\varepsilon/2\right)}\,.\label{eq:kappa_ren-minus-kappa}
\end{eqnarray}
It is clear that the FRG result should be compared against this
expression since $\kappa$ is negligible in the inertial range. 
The expression (\ref{eq:kappa_ren-minus-kappa}) exhibits three main features:
it is proportional to $D_{0}$, it diverges as $1/\varepsilon$ as $\varepsilon\rightarrow0$,
and it is proportional to $m^{-\varepsilon}$ (which signals an IR
singularity as $m\rightarrow0$). All these three aspects are shared
by the coefficient appearing in the FRG formula (\ref{eq:dimless-FP-sol-Gqq-4}).

Now, even neglecting the molecular viscosity, the two coefficients share many
similarities but are not equal, which is to be expected, since the setting and assumptions
 are different. For instance the coefficient (\ref{eq:kappa_ren-minus-kappa})
is IR divergent if no IR mass/regulator is added, which renders
 difficult a meaningful comparison beyond the qualitative features
outlined above. However, note that if one chooses a mass cutoff
 kernel defined by $S_{{\rm vel}}+\Delta S_{k}=\int\frac{1}{2}v\left(-\partial^{2}+k^{2}\right)^{\frac{d+\varepsilon}{2}}v$
and eventually sets $k=m$, one finds that the exponent in (\ref{eq:dimless-FP-sol-Gqq-4})
has the same expression as in (\ref{eq:kappa_ren-minus-kappa}).
In principle, one can reconstruct the correlations of any microscopic theory by
solving the FRG equation exactly,  strictly taking the limit $k\rightarrow 0$, and keeping $k\neq m$.\\

Let us now make two additional remarks.
 First, the same framework of the Dyson equations can be used
  to determine the behavior of the spectrum in the  mildly non-linear regime,
   that is in the dissipative range, but close to the inertial range where the convection still plays a role.
    We find that the energy spectrum in this range exhibits an intermediate power law regime following $p^{-(d+2+\varepsilon)}$,
 which is shown in Appendix~\ref{app:dissipative}.

Finally, let us comment on the discretization scheme
 of the  SDE
(\ref{eq:Kraichnan-model_SDE}). 
This equation
 is written in the Stratonovich convention. 
In the Ito convention, it reads (see, e.g., Ref.~\cite{Kupiainen:2006em})
\begin{eqnarray}
\partial_{t}\theta +v^{i} \partial_{i}\theta 
-\left(\frac{\kappa}{2}\delta^{ij}+\frac{1}{2}D^{ij}\left(0\right)\right)\partial_{i}\partial_{j}\theta
& = & f \,,\label{eq:Kraichnan-model-Ito-SDE-1}
\end{eqnarray}
where $D^{ij}\left(\vx-\vy\right)$ denotes the spatial part of the velocity
covariance (\ref{eq:vel-2pt-correlation-function}), explicitly
$\langle v^{i}\left(t,\vx\right)v^{j}\left(t^{\prime},\vy\right)\rangle=\delta\left(t-t^{\prime}\right)D^{ij}\left(\vx-\vy\right)$.
 The new term in the SDE can be expressed as
\begin{eqnarray*}
\frac{1}{2}D^{ij}\left(0\right) & = & \frac{1}{2}D_{0}\int_{\vq}\frac{P_{ij}(\vq)}{\left(q^{2}+m^{2}\right)^{\frac{d}{2}+\frac{\varepsilon}{2}}} \nonumber \\
&=&
D_{0}\frac{\left(d-1\right)}{2d}\frac{\Gamma\left(\varepsilon/2\right)m^{-\varepsilon}}{\left(4\pi\right)^{d/2}\Gamma\left(d/2+\varepsilon/2\right)}\delta_{ij}\, .
\end{eqnarray*}
Hence, one observes that the coefficient in front of $\delta_{ij}$ is exactly
the quantity appearing in (\ref{eq:kappa_ren-minus-kappa}). It follows
that Eq.~(\ref{eq:Kraichnan-model-Ito-SDE-1}) can be written
as
\begin{eqnarray}
\partial_{t}\theta\left(t,\vx\right)+v^{i}\left(t,\vx\right)\partial_{i}\theta\left(t,\vx\right)-\frac{\kappa_{{\rm ren}}}{2}\partial^{2}\theta\left(t,\vx\right) & = & f\left(t,\vx\right)\,. \quad
\label{eq:Kraichnan-model-Ito-SDE-2}
\end{eqnarray}
From the SDE (\ref{eq:Kraichnan-model-Ito-SDE-2}), one
can estimate the temporal dependence of $\bar{G}_{\theta\theta}\left(t,\vp\right)$
via the following argument. Let us neglect the convection and the
forcing term. In this approximation
 $\partial_{t}\langle\theta\left(t,\vx\right)\theta\left(0,\vy\right)\rangle=\frac{\kappa_{{\rm ren}}}{2}\partial_{x}^{2}\langle\theta\left(t,x\right)\theta\left(0,y\right)\rangle$,
which leads precisely to the temporal behavior found via field theoretical
methods, namely $\bar{G}_{\theta\theta}\left(t,\vp\right)\propto e^{-\frac{\kappa_{{\rm ren}}}{2}p^{2}\left|t\right|}$.
Summarizing, the effect of computing $\Sigma_{\theta\bar{\theta}}$
(or, equivalently, to renormalize the microscopic molecular diffusivity)
 simply amounts to introduce an ``Ito correction'' to the molecular
diffusivity. In this sense, the Ito convention makes the mixing between
the velocity and the scalar more transparent.
A more refined approach has been developed in  Ref.~\onlinecite{MitraPandit2005}, 
where a differential equation for the two-point correlation function
at unequal time was found and investigated numerically confirming
the exponential decay of the two-point correlation function.

\section{Closed flow equation for the correlation functions of advected
scalars} \label{sec:closed-FRGE-for-advected-scalars}

In the Kraichnan model, the temporal dependence of the two-point correlation function
of the scalar can be obtained from various approaches, as shown in  
 Sec.~\ref{subsec:2pt-correlation-function-from-Dyson-eq}. This is due 
 to the  key simplifying feature of this model, which is the white-in-time
Gaussian statistics of the velocity. However, the resulting temporal dependence
for the scalar is far from that of scalars in actual flows. However, the approximation
scheme developed in Secs.~\ref{subsec:Closed-flow-equations} and
\ref{subsec:2pt-function-in-large-P} is not based on the white-in-time
statistics of the velocity  but on the symmetries of the action functional. 
As we shall show in Sec.~\ref{subsec:Action-and-symmetries-for-NS_plus_scalar},
the symmetries for scalars advected by a velocity field satisfying
the NS equation are not much different from the symmetries of the Kraichnan model, 
 such that the same scheme is also applicable to real scalars, contrarily to the perturbative
  approach. The main objective of this section is thus to derive
  and solve the closed flow equation for the two-point correlation
 function of the scalar field advected by NS turbulence by exploiting the strategy already followed
 for the Kraichnan model.

\subsection{Action and symmetries for scalars advected by a NS velocity \label{subsec:Action-and-symmetries-for-NS_plus_scalar}}

To obtain the action functional for the scalar field advected by a NS turbulent flow, 
we apply the MSRJD procedure to the system of equations
composed by the Navier-Stokes equation for an incompressible fluid and by the advection-diffusion equation (\ref{eq:Kraichnan-model_SDE}) for the scalar. In order to describe a stationary
turbulent state, we consider the  NS equation equipped with a
Gaussian stochastic forcing. Following he conventions adopted in  Ref.~\onlinecite{Canet:2014dta},
the  action functional associated to the
NS equation is given by
\begin{eqnarray}
S_{{\rm NS}} & = & \int_{t\vx}\left[\bar{v}^{i}\left(\partial_{t}+v^{j}\partial_{j}-\kappa_{\nu}\partial^{2}\right)v^{i}+\bar{v}^{i}\partial_{i}p+\bar{p}\partial_{i}v^{i}\right]\nonumber\\
&-&\int_{t\vx\vx'}\frac{1}{2}\bar{v}^{i}N_{k,ij}\bar{v}^{j}\,,
 \label{eq:action-NS}
\end{eqnarray}
where $\kappa_{\nu}$ denotes the kinematic viscosity (the notation $\nu$ is avoided to prevent confusion with the velocity $v$), $N_{k}$ 
is the covariance of the stochastic forcing of the fluid velocity
 in the NS equation, for which the integral scale $L$ of the fluid
 is identified with the inverse of the RG scale $k$,
$p$ is the pressure, and $\bar{p}$ serves
as a Lagrange multiplier which ensures the velocity to be divergenceless.
The pressure sector is fully non-renormalized, as can be shown by considering
gauged shifts in $p$ and $\bar{p}$, see   Ref.~\onlinecite{Canet:2014dta} for the explicit Ward identities.
 The covariance $N_{k,ij}$ can be chosen diagonal in components $N_{k,ij}=\delta_{ij}N_k$ because of incompressibility,
 and $N_k$ is taken  of the same form as the
 one of the scalar forcing $M_{k}$ defined in (\ref{eq:def-forcing-M_k}).

The total action of the NS velocity plus scalar system is thus given by
\begin{eqnarray}
S & = & \int_{t\vx}\left[\bar{\theta}\left(\partial_{t}+v^{i}\partial_{i}-\frac{\kappa}{2}\partial^{2}\right)\theta\right]-\int_{t\vx\vx'}\frac{1}{2}\bar{\theta}M_{k}\bar{\theta}+S_{{\rm NS}}\,.\nonumber\\
\label{eq:action-NS_plus_scalar}
\end{eqnarray}
with the ghost sector implicit. 
The action (\ref{eq:action-NS_plus_scalar}) shares some similarity with the Kraichnan
 action (\ref{eq:Kraichnan-model-action}) with the main difference
that now the velocity does not appear quadratically but in the more
complicated form $S_{{\rm NS}}$. However, the symmetries
are essentially unaltered, as we now discuss. 

In the FRG context, the symmetries and extended symmetries of the NS action (\ref{eq:action-NS})
have been discussed in  Ref.~\onlinecite{Canet:2014dta}, to which we refer for
a detailed presentation. Here we limit ourselves to discussing the most
relevant points and the extension to the action (\ref{eq:action-NS_plus_scalar}).\\

\emph{Time-gauged Galilei transformation}. The Galilean transformation
 has the same expression as (\ref{eq:Galilei-transformation}) with the response velocity field
  and pressure fields transforming as scalars 
   $$ \bar{\varphi}\left(t,\vx\right)  \rightarrow  \varphi\left(t,\vx\right)-\varepsilon^{i}\left(t\right)\partial_{i}\varphi\left(t,\vx\right) $$
for $\varphi=\bar{v}^{j}, p$ or $\bar{p}$.
The associated Ward identity reads
\begin{eqnarray}
&0&  =  
\int_{\vx}\partial_{t}^{2}\bar{v}^{i}\left(t,\vx\right)
+\int_{\vx} \Biggr[\partial_{t}\frac{\delta\Gamma_{k}\left[\varphi\right]}{\delta v^{i}\left(t,\vx\right)}+\frac{\delta\Gamma_{k}\left[\varphi\right]}{\delta v^{j}\left(t,\vx\right)}\partial_{i}v^{j}\left(t,\vx\right)
\nonumber \\
 &  & +\frac{\delta\Gamma_{k}\left[\varphi\right]}{\delta\bar{\theta}\left(t,\vx\right)}
 \partial_{i}\bar{\theta}\left(t,\vx\right)+\frac{\delta\Gamma_{k}\left[\varphi\right]}{\delta\theta\left(t,\vx\right)}\partial_{i}\theta\left(t,\vx\right) +\frac{\delta\Gamma_{k}\left[\varphi\right]}{\delta\bar{p}\left(t,\vx\right)}
 \partial_{i}\bar{p}\left(t,\vx\right)\nonumber \\
 &  & +\frac{\delta\Gamma_{k}\left[\varphi\right]}{\delta p\left(t,\vx\right)}\partial_{i} p\left(t,\vx\right) +\frac{\delta\Gamma_{k}\left[\varphi\right]}{\delta\bar{v}^{j}\left(t,\vx\right)}\partial_{i}\bar{v}^{j}\left(t,\vx\right)
 \Biggr]\,. \qquad
\label{eq:WI-T-gauged-Galiei-for-NS_plus_scalar}
\end{eqnarray}
As in the case of the Kraichnan model, the Ward identity for time-gauged
Galilean invariance is the crucial ingredient for the closure of the
flow equation of any $n$-point correlation function in the limit of large wavenumber
 since it allows one to express a $\left(n+1\right)$-vertex
 with one velocity wavevector set to zero in terms of $n$-point vertices.\\

\emph{Time-gauged shifts of the response fields.} An important extended symmetry of the NS
 action is related to the following time-gauged transformation
$\delta\bar{v}^{i}\left(t,\vx\right)=\bar{\varepsilon}^{i}\left(t\right)$
and $\delta\bar{p}\left(t,\vx\right)=\bar{\varepsilon}^{i}\left(t\right){v}^{i}\left(t,\vx\right)$.
It yields the Ward identity
\begin{eqnarray*}
\int_{\vx}\frac{\delta\Gamma_{k}}{\delta\bar{v}_{i}\left(t,\vx\right)} & = & \int_{\vx}\partial_{t}v^{i}+\int_{\vx}v^{j}\left(t,\vx\right)\partial_{j}v^{i}\left(t,\vx\right)\,,
\end{eqnarray*}
which entails the non-renormalization of the Lagrangian time derivative 
 term $\int\bar{v}\left(\partial_{t}+v\cdot\partial\right)v$
in the bare action, and the vanishing of all the vertices of $n>2$ points with 
 one zero wavevector carried by a response velocity. \\

\emph{Shift symmetry of the scalar fields.}
The time-gauged shift symmetry $\theta\left(t,\vx\right)\rightarrow\theta\left(t,\vx\right)+\epsilon\left(t\right)$ is the same as in the Kraichnan model and
 the Ward identity (\ref{eq:WI-t-local-shift-1}) is also valid for the NS scalar.
  Thus, any $\left(n>2\right)$-point vertex with one zero wavevector carried by a
 scalar field  vanishes. 
 
 For the response scalar field, the time-gauged shift now involves the coupled transformation $\bar{\theta}\left(t,\vx\right)\rightarrow\bar{\theta}\left(t,\vx\right)+\varepsilon\left(t\right)$
and $\bar{p}\left(t,\vx\right)\rightarrow\bar{p}\left(t,\vx\right)+\varepsilon\left(t\right)\theta\left(t,\vx\right)$, which leads to the modified Ward identity
\begin{eqnarray*}
0 & = & -\int_{\vx}\left[\partial_{t}\theta\left(t,\vx\right)+v^{i}\left(t,\vx\right)\partial_{i}\theta\left(t,\vx\right)\right]+\int_{\vx}\frac{\delta\Gamma\left[\varphi\right]}{\delta\bar{\theta}\left(t,\vx\right)}\,.
\end{eqnarray*}
This identity entails that the term $\int \bar\theta v_j\p_j \theta$ is not renormalized. 
Furthermore, it yields that  any vertex with a wavevector carried by a $\bar{\theta}$ set
to zero is actually given by its bare expression, and thus vanishes when $n>2$.\\

\emph{Synthesis}. 
The key point is that the Ward identities for the scalar advected by the NS flow, although 
 slightly modified compared to the ones of the Kraichnan scalar, imply the same
 consequence for 1-PI vertices, which is the cornerstone of the large wavenumber closure:
  if at least one wavevector of the vertex is set to zero, then it vanishes 
  except when  the wavevector is carried
 by a velocity field, in which case it is controlled by (\ref{eq:WI-T-gauged-Galiei-for-NS_plus_scalar}).\footnote{Possible contributions to the flow equation from any bare-like three-point
vertex actually vanishes by incompressibility.}

\subsection{Anomalous dimensions}

In this section, we discuss the constraints on the anomalous
 dimensions  imposed
by the symmetries analyzed in Sec.~\ref{subsec:Action-and-symmetries-for-NS_plus_scalar}.
The analysis proceeds along the same line as for the  Kraichnan model in 
Sec.~\ref{subsec:Non-renormalization-theorem-and-anomalous-dim}, and the outcome
 is similar in that the anomalous dimensions are essentially fixed by the symmetries
  and the requirement of stationary turbulence.

We parametrize the EEA for the scalar advected by the NS flow in a similar way as the one for the Kraichnan scalar
\begin{eqnarray*}
\Gamma_{k} & = & \int_{t\vx}\left[Z_{k,\bar{v}}^{1/2}\bar{v}^{i}\left(\partial_{t}+Z_{k,v}^{1/2}\lambda_{k,v}v^{j}\partial_{j}-{\kappa_{k,\nu}}\partial^{2}\right)Z_{k,v}^{1/2}v^{i}\right]\\
 &  & +\int_{t\vx}\left[Z_{k,\bar{\theta}}^{1/2}\bar{\theta}\left(\partial_{t}+Z_{k,v}^{1/2}\lambda_{k,\theta}v^{i}\partial_{i}-\frac{\kappa_{k}}{2}\partial^{2}\right)Z_{k,\theta}^{1/2}\theta\right]+\cdots\,,
\end{eqnarray*}
where we introduced the field renormalizations $Z_{k,\varphi^{i}}$,
the couplings $\lambda_{k,v}$ and $\lambda_{k,\theta}$, and the running
viscosity and running molecular diffusivity, $\kappa_{k,\nu}$ and $\kappa_{k}$
respectively. In full analogy with the analysis of section \ref{subsec:Non-renormalization-theorem-and-anomalous-dim},
one deduces  from the non-renormalization theorems the following constraints: $Z_{k,\bar{v}}^{1/2}Z_{k,v}^{1/2}=1$,
$Z_{k,v}^{1/2}\lambda_{k,v}=1$, $Z_{k,\bar{\theta}}^{1/2}Z_{k,\theta}^{1/2}=1$,
and $Z_{k,v}^{1/2}\lambda_{k,\theta}=1$. In terms of couplings and anomalous
dimensions, it follows that $\eta_{\bar{v}}=-\eta_{v}$, $\eta_{\bar{\theta}}=-\eta_{\theta}$,
and $\lambda_{k,\theta}=\lambda_{k,v}\equiv \lambda_k=1$, which plays the role of the coupling
appearing in the Galilean covariant derivative. We also associate
 anomalous dimensions $\eta_{\kappa_{\nu}}$ and $\eta_{\kappa}$ to the viscosity
and the molecular diffusivity, respectively.

We assume that the scalar is passive, which means that its backreaction onto
the velocity flow is negligible. In the framework of the RG, this
translates into the statement that none of the quantities related
to the velocities, in particular $\eta_{v}$, $\eta_{\kappa_{\nu}}$, and $\lambda_{k,v}$,
are affected by the scalar field and its fluctuations. Therefore,
in order to determine such quantities, we  simply neglect the
terms in the EAA which contains a scalar field. This part of
the EAA then simply corresponds to the NS equations,  already studied in  Ref.~\onlinecite{Canet:2014dta},
 and of which we only give a concise account. In essence, a  
dimensionless coupling $\hat{\lambda}_{k,v}$ is introduced as  
$\lambda_{k,v}\equiv\hat{\lambda}_{k,v}{\kappa_{k,\nu}}^{3/2}k^{-d/2+1}$
 and its  beta function  is simply given by $\partial_{s}\hat{\lambda}_{k,v}=\left(3\eta_{\kappa_{\nu}}-\eta_{\bar{v}}+d-2\right)\hat{\lambda}_{k,v}/2$
  since $\lambda_{k,v}$ is not renormalized.
It follows that  any non-Gaussian fixed point satisfies $3\eta_{\kappa_{\nu}}-\eta_{\bar{v}}+d-2=0$.
Finally, in analogy with the arguments of Sec.~\ref{subsec:Non-renormalization-theorem-and-anomalous-dim},
one may determine the anomalous dimensions by  requiring $\lim_{k\rightarrow0}\langle f\cdot v\rangle$
 and $\lim_{\kappa_\nu\rightarrow0}{\kappa_\nu}\langle\partial_{i}v_{j}\partial_{i}v_{j}\rangle$
to be non-zero constants. This leads in $d=3$ to fix  $\eta_{\bar{v}}=d+2=5$,
 which then yields that at a non-Gaussian fixed point $\eta_{\kappa_\nu}=4/3$.

Let us now turn to the anomalous dimensions and the couplings inherent to
the scalar dynamics. We introduce the dimensionless coupling $\hat{\kappa}_k\equiv\kappa_{k}/{\kappa_{k,\nu}}$.
The associated beta function is given by  $\partial_{s}\hat{\kappa}_k=\left(-\eta_{\kappa}+\eta_{\kappa_\nu}\right)\hat{\kappa}_k$.
At any non-trivial fixed point of $\hat{\kappa}$, one thus has $\eta_{\kappa}=\eta_{\kappa_\nu}$.
 This means that the dynamical critical
exponent $z$ is controlled by the viscosity and that it also 
 applies to the scalar sector. To determine the anomalous dimensions
of the scalars, we proceed as in Sec.~\ref{subsec:Non-renormalization-theorem-and-anomalous-dim}.
One may either require that $\lim_{k\rightarrow0}\langle f_{\theta}\theta\rangle$
approaches a constant, which implies $\eta_{\bar{\theta}}=d+2$, or 
that $\lim_{\kappa\rightarrow0}\kappa\langle\partial_{i}\theta\partial_{i}\theta\rangle$
is finite, which leads to the same result. In this latter case, the difference with respect to 
Eq.~(\ref{eq:dissipation-limit-kappa_to_zero-and-kappa-power}) is that
 the dissipation scale is provided by $\eta_{{\rm diss}}^{-1}\equiv\left(\epsilon_{\theta}/\kappa^{3}\right)^{1/4}$,
with $\epsilon_{\theta}$  the scalar energy density conserved
through the cascade. 

\subsection{Closed flow equation for the two-point correlation function}

Since the general form of the flow equation and the structure of the Ward
identities for the passive scalar in the NS flow are the same as in the Kraichnan
model, one can closely follow the derivation of Secs.~\ref{subsec:Closed-flow-equations}
and \ref{subsec:2pt-function-in-large-P} to obtain a closed flow equation for any $n$-point correlation
 function of the scalar field in the limit of large wavenumber. 

The inverse propagator $\big(\Gamma_k^{(2)}+{R}_k\big)$ is now a $8\times 8$ matrix,
 comprising the scalar field sector ($\theta,\bar{\theta},c,\bar{c}$) and the velocity
  sector ($v_i,\bar{v}_i,p,\bar{p}$), but it has a simple structure since it is block-diagonal,
   and also diagonal in frequency and wavevector in Fourier space.
    It follows that the renormalized propagator bares a similar structure as Eq.~(\ref{eq:propag}), with
 the $G_{vv}$  now replaced by a $4\times 4$ matrix whose general form can be found in Ref.~\onlinecite{Canet:2014dta}.
  One can show that the pressure sector decouples (as the ghost sector) and the only part which plays a role is the velocity
   sector, whose propagator is given by
\begin{eqnarray*}
\left(\bar{\Gamma}_{k}^{\left(2\right)}+R_{k}\right)^{-1}_{v\bar{v}} & = & \left(\begin{array}{cc}
\bar{G}_{v^\alpha v^\beta} & \bar{G}_{v^\alpha\bar{v}^\beta} \\
\bar{G}_{\bar{v}^\alpha v^\beta} & 0 
\end{array}\right)\,,
\end{eqnarray*}
with 
\begin{eqnarray*}
\bar{G}_{v^\alpha \bar{v}^\beta}( \omega,\vq) &=& 
\frac{P_{\alpha\beta}(\vq)}{\bar{\Gamma}^{(2)}_{v^\alpha \bar{v}^\beta}( -\omega,\vq) + R_{k,v\bar{v}}(\vq)}\,, \\
\bar{G}_{v^\alpha {v}^\beta}( \omega,\vq)&=& -P_{\alpha\beta}(\vq) 
\frac{\bar{\Gamma}^{(2)}_{\bar{v}^\alpha \bar{v}^\beta}( \omega,\vq) -2 N_{k,\alpha\beta}(\vq)}{\left| \bar{\Gamma}^{(2)}_{v^\alpha \bar{v}^\beta}+R_{k,v\bar{v}}(\vq)\right|^2},
\end{eqnarray*}
where the field functional derivative are indicated as indices on the $\Gamma^{(2)}$ to alleviate notation.

We now consider the generic flow equation (\ref{eq:FRGE-W(n)-1}) for a $n$-point correlation function $W_{k}^{\left(n\right)}$.
In the large wavenumber limit, since we have shown that all vertices with one zero-wavevector
 vanish but the ones for which it is carried by a velocity field, the only remaining term in
  this flow equation is
\begin{eqnarray*}
&&
\partial_{s}\frac{\delta^{n}W_{k}\left[J\right]}{\delta J^{1}\cdots\delta J^{n}} 
= -\frac{1}{2} H_{k,i j}  \frac{\delta}{\delta v^{i}}\frac{\delta}{\delta v^{j}}\frac{\delta^{n}W_{k}\left[J\right]}{\delta J^{1}\cdots\delta J^{n}}\,,
\end{eqnarray*}
where we introduced the notation
\begin{eqnarray*}
&&
H_{k,i j}\left(\omega,\vq\right) \\
&& \qquad\qquad
\equiv \left(2\bar{G}_{v^{i}v^{\alpha}}\partial_{s}R_{k,v^{\alpha}\bar{v}^{\beta}} \Re (\bar{G}_{\bar{v}^{\beta}v^{j}})+\bar{G}_{v^{i}\bar{v}^{\alpha}}\partial_{s}N_{k,\alpha\beta}\bar{G}_{\bar{v}^{\beta}v^{j}}\right)\,.
\end{eqnarray*}
Hence this equation can be closed exploiting the time-gauged Galilean Ward identity 
\begin{eqnarray*}
&& 
\frac{\delta}{\delta v^{i}\left(-\omega,-\vq\right)}\frac{\delta}{\delta v^{j}\left(\omega,\vq\right)}W_{k}^{\left(n\right)}\left(\omega_{1},\vp_{1},\cdots\right)\Bigr|_{\vq=0} \\
&& \qquad\qquad\qquad\qquad
= {\cal D}^{i}\left(-\omega\right){\cal D}^{j}\left(\omega\right)W_{k}^{\left(n\right)}\left(\omega_{1},\vp_{1},\cdots\right)\,.
\end{eqnarray*}
One thus also sobtains  for the scalar in a turbulent NS flow a closed flow equation for any $n$-point correlation
 function of any fields since at this stage, the $J^m$ are any of the sources.

In the following, we focus on the two-point correlation function of the scalar field. The explicit expression 
 of its flow equation, with an inverse Fourier transform to revert to time-wavevector coordinates, is given at large $p$ by 
\begin{eqnarray}
&&
\partial_{s}\bar{G}_{\theta\theta}\left(t,\vp\right) \nonumber\\
&&= \frac{1}{2}\int_{\omega \vq} H_{k,ij }\left(\omega,\vq\right)
p^{i}p^{j }\Biggr\{\frac{2-2\cos\left(t\omega\right)}{\omega^{2}}\Biggr\}\bar{G}_{\theta\theta}\left(t,\vp\right)\,.\label{eq:FRGE-Gqq(tp)-advected-scalar-1}
\end{eqnarray}
Since $H_{k,ij }= H_k P_{ij }$, one can further simplify 
 Eq.~(\ref{eq:FRGE-Gqq(tp)-advected-scalar-1}) replacing
 $H_{k,ij } p^i p^j  \to \frac{d-1}{d} H_k p^2$ in the integral using isotropy.
Eq.~(\ref{eq:FRGE-Gqq(tp)-advected-scalar-1}) is formally very
similar to Eq.~(\ref{eq:FP-FRGE-Gqq-1}) obtained for the
Kraichnan scalar. However, there are some crucial differences.
 First, in the case of the NS passive scalar, the velocity
propagators are renormalized and are not known exactly. Second,
the velocity propagators, which enter in (\ref{eq:FRGE-Gqq(tp)-advected-scalar-1}) through $H_{k}$,
have a non-trivial frequency dependence, contrarily to  the 
 case of the Kraichnan model. Since their exact form is not known,
 one cannot obtain an explicit expression for the resulting
frequency integral. However, these integrals can be simplified in both the short-time and long-time limits, 
 which we study below.

\subsubsection{Short-time limit of the two-point correlation function}

Let us consider the limit $t\rightarrow0$ of Eq.~(\ref{eq:FRGE-Gqq(tp)-advected-scalar-1}).
In this limit,  the cosine can be expanded which yields
\begin{eqnarray*}
\partial_{s}\bar{G}_{\theta\theta}\left(t,\vp\right) & = & \frac{d-1}{2d}\int_{\omega \vq}H_{k}\left(\omega,\vq\right)p^2 t^{2}\bar{G}_{\theta\theta}\left(t,\vp\right)\,,
\end{eqnarray*}
 since  the frequency integration converges
thanks to the function $H_{k}$.
The dimensionless two-point function is defined by $\bar{G}_{\theta\theta}\left(t,\vp\right)\equiv k^{\eta_{\kappa}-\eta_{\bar{\theta}}}\hat{\bar{G}}_{\theta\theta}\left(\hat{t},\hat{\vp}\right)$,
and it satisfies the flow equation
\begin{eqnarray*}
&&
\left(\partial_{s}+\left(\eta_{\kappa}-\eta_{\bar{\theta}}\right)+\left(2-\eta_{\kappa}\right)\hat{t}\partial_{\hat{t}}-\hat{\vp}\partial_{\hat{\vp}}\right)\hat{\bar{G}}_{\theta\theta}
\left(\hat{t},\hat{\vp}\right) \\
&& \qquad\qquad\qquad\qquad
= \hat{\alpha}_k \hat{\lambda}_k^{2}\hat{p}^{2}\hat{t}^{2}\hat{\bar{G}}_{\theta\theta}\left(\hat{t},\hat{\vp}\right)\,, 
\end{eqnarray*}
where we introduced the scale dependent parameter $\hat{\alpha}_k\equiv\frac{d-1}{2d}\int_{\hat{\omega}\hat{\vq}}\hat{H}_k\left(\hat{\omega},\hat{\vq}\right)$.
At the fixed point, this parameter is just a number $\hat{\alpha}_k\to \hat{\alpha}_{{\rm s}}$
 and the fixed-point equation reads, now specifying to $d=3$
\begin{eqnarray}
&&
\left(-\frac{11}{3}+\frac{2}{3}\hat{t}\partial_{\hat{t}}-\hat{\vp}\partial_{\hat{\vp}}\right)\hat{\bar{G}}_{\theta\theta}\left(\hat{t},\hat{\vp}\right) 
\nonumber \\
&& \qquad\qquad\qquad\qquad\qquad 
= \hat{\alpha}_{{\rm s}}\hat{\lambda}^{2}\hat{p}^{2}\hat{t}^{2}\hat{\bar{G}}_{\theta\theta}\left(\hat{t},\hat{\vp}\right)\,.\label{eq:FP-FRGE-Gqq-short-time}
\end{eqnarray}
The solution of this equation is
\begin{eqnarray*}
\hat{\bar{G}}_{\theta\theta}\left(\hat{t},\hat{\vp}\right) & = & \hat{p}^{-\frac{11}{3}}e^{-\frac{3}{2}\hat{\alpha}_{{\rm s}}\hat{\lambda}^{2}\hat{p}^{2}\hat{t}^{2}}\hat{f}\left(\hat{p}^{2/3}\hat{t}\right)\,,
\end{eqnarray*}
where $\hat{f}\left(\hat{p}^{2/3}\hat{t}\right)$ is a universal scaling function, which could be computed by explicitly
 integrating the flow from the initial condition.
Reverting to dimensionful variables, and neglecting the subleading time dependence of the scaling function, one obtains
\begin{eqnarray}
G_{\theta\theta}\left(t,\vec{p}\right) = C_{\rm s} \epsilon_\theta \epsilon^{-1/3}_v 
{p}^{-\frac{11}{3}} e^{-\hat{\alpha}_{\rm s} (L/\tau_0)^2 p^2 t^2}\,,
\quad
\label{eq:2pt-correlation-short-time}
\end{eqnarray}
where $\tau_0\equiv \left(L^2/\epsilon_v \right)^{-1/3}$ denotes the eddy-turnover time at the energy injection scale
and $C_{\rm s}$ and $\hat{\alpha}_{\rm s}$ are non-universal constants (the factor 3/2 was absorbed in
 the latter).
In Eq.~(\ref{eq:2pt-correlation-short-time}) $\epsilon_\theta$ and $\epsilon_v$ 
denote the mean rate of energy injection of the scalar and of the velocity,
respectively.

\subsubsection{Large-time limit of the two-point correlation function}

We now consider the limit $t\rightarrow\infty$ of Eq.~(\ref{eq:FRGE-Gqq(tp)-advected-scalar-1}).
In this limit, it is straighforward to check that the frequency integration
is dominated by the term in the curly brackets of Eq.~(\ref{eq:FRGE-Gqq(tp)-advected-scalar-1}).
 Performing the frequency integration, one obtains 
\begin{eqnarray*}
\partial_{s}\bar{G}_{\theta\theta}\left(t,\vp\right) & = & \frac{d-1}{2d}\int_{\vq}H_{k}\left(0,\vq\right)p^{2}\left|t\right|\bar{G}_{\theta\theta}\left(t,\vp\right)\,.
\end{eqnarray*}
In  analogy with the short-time limit, we thus obtain the
fixed point equation in $d=3$
\begin{eqnarray}
&&
\left(-\frac{11}{3}+\frac{2}{3}\hat{t}\partial_{\hat{t}}-\hat{\vp}\partial_{\hat{\vp}}\right)\hat{\bar{G}}_{\theta\theta}\left(\hat{t},\hat{\vp}\right) 
\nonumber \\
&& \qquad\qquad\qquad\qquad\qquad\quad
= \hat{\alpha}_{{\rm \ell}}\hat{\lambda}^{2}\hat{p}^{2}\left|\hat{t}\right|\hat{\bar{G}}_{\theta\theta}\left(\hat{t},\hat{p}\right)\,.\label{eq:FP-FRGE-Gqq-large-time}
\end{eqnarray}
where $\hat{\alpha}_{{\rm \ell}}$ is the fixed point value of $\frac{1}{3}\int_{\hat{\vq}}\hat{H}_k\left(0,\hat{\vq}\right)$.
The solution to equation (\ref{eq:FP-FRGE-Gqq-large-time}) is given by
\begin{eqnarray*}
\hat{\bar{G}}_{\theta\theta}\left(\hat{t},\hat{\vp}\right) & = & \hat{p}^{-\frac{11}{3}}e^{-\frac{3}{4}\hat{\alpha}_{{\rm \ell}}\hat{\lambda}^{2}\hat{p}^{2}|\hat{t}|}\hat{f}\left(\hat{p}^{2/3}\hat{t}\right)\,,
\end{eqnarray*}
which in dimensionful variables implies
\begin{eqnarray}
G_{\theta\theta}\left(t,\vec{p}\right) = C_{\rm \ell} \epsilon_\theta \epsilon^{-1/3} 
{p}^{-\frac{11}{3}} e^{-\hat{\alpha}_{\rm \ell} \frac{L^2}{\tau_0} p^2 |t|}\,,
\label{eq:2pt-correlation-long-time}
\end{eqnarray}
where $C_{\rm \ell}$ and $\hat{\alpha}_{\rm \ell}$ are non-universal constants (with the numerical factor absorbed in the latter).

 Equations (\ref{eq:2pt-correlation-short-time}) and (\ref{eq:2pt-correlation-long-time}) constitute the main results of this work
 and generalize the temporal dependence found for the Kraichnan model
to scalar fields advected by a NS flow. As anticipated, when the carrier turbulent velocity field
  have some temporal correlations, which is always the case for any realistic flows, then the short-time behavior
   of the scalar is Gaussian in $t$. The exponential decay is only observed at sufficiently large time scales.
 This is the main difference with the scalar advected by the idealized Kraichnan stochastic velocity,
  where the short-time regime is eliminated by the time delta-correlation of the velocity covariance.

\section{Summary and perspectives} \label{sec:summary}

In this work, we have studied the temporal dependence
 of the two-point correlation function 
of turbulent passive scalar fields in the inertial-convective range. We analyzed
the symmetries, and extended symmetries,  of the action functional of the system,
 and derived the corresponding Ward identities. These identities
  are the crucial ingredients which enabled us to obtain, within the FRG framework in the limit of large wavenumbers,  a closed flow equation for
   generic Eulerian $n$-point correlation functions of the scalar field, advected both by the Gaussian stochastic velocity
    of the Kraichnan model, or by a turbulent NS velocity field.
   
    We have focused on the solution at the fixed point of the two-point
     correlation function of the scalar. We have shown that
      its temporal decay exhibits two regimes: a Gaussian decay at small time delays,
       which crosses over to an exponential decay at large time delays, 
        with a coefficient proportional to $k^{2}$ in both regimes.
        This time dependence explicitly breaks standard scale invariance.
       This demonstrates that the scalar field, in the inertial-convective range considered here,
        inherits the time correlations of the carrier fluid, since this exactly
         corresponds to the behavior of a  NS velocity
         field \cite{Tarpin:2018yvs}.
 
 The Kraichnan model replaces the carrier NS fluid velocity by a Gaussian
  stochastic field with delta-correlation in time.
   This leads to great simplifications, as was already recognized in 
    many works. As a consequence, we obtained an explicit expression for the prefactor
     in the exponential, and we also derived the temporal behavior
     of the two-point function in this model using a perturbative approach,
      which is not possible for the NS flow. This approach yields
       results compatible with the FRG approach, and provides a useful connection with former results.
     The prize of this simplification is that it significantly alters
   the temporal behavior of the advected scalar, since the small-time
    regime is destroyed in this case.

  The approximation scheme used in this work, based on a large wavenumber expansion, is rooted in the symmetries of
the system. Since the symmetries of the advected scalar field are also
present  in other regimes, say the viscous-convective regime,
it is conceivable that this approach can be extended
 outside the inertial-convective range. We leave this task for future
work.  Let us emphasize that all the results obtained in this work 
 on  the temporal behavior of turbulent passive scalar fields
 can be tested in direct numerical simulations, which is underway \cite{Gorbunova:2021b}.
 These results may also initiate experimental investigations.	
 
 Another important direction which deserves further study is the combined use
 of numerical solutions to the FRG flow equations, via ans{\"a}tze, and the symmetry
 arguments presented in this paper. This will allow one in particular to 
  study  correlation functions which include
 composite operators, and investigate their anomalous scalings. \footnote{We refer to 
 Ref.~\onlinecite{Pagani:2016pad,Pagani:2017tdr,Pagani:2020ejb,Rose:2015bma} 
for details regarding the composite operators within the FRG framework.} 
 We hope to tackle  these issues in the near future.

\begin{acknowledgments}
The authors thank B.~Delamotte and M.~Tarpin for discussions and
collaboration at the initial stages of the project, 
 A.~Gorbunova for discussions and collaboration on
related projects,
and G.~Eyink for useful references.
L.C.~acknowledges support from the ANR-18-CE92-0019 Grant NeqFluids and support from Institut Universitaire de France.
\end{acknowledgments}

\appendix

\section{Yaglom relation from the path integral} \label{app:Yaglom_rel}

After the celebrated derivation by Kolmogorov of the exact formula for the third-order structure
function from the Navier-Stokes equations, Yaglom established
the  analogous formula  for scalar turbulence \cite{Yaglom49}.
 It was shown in~ Ref.~\onlinecite{Canet:2014cta} that the K\'arman-Howarth relation
  on which is based Kolmogorov's exact identity for the third-order
    structure function can be retrieved within a path integral formalism
 by considering a gauged shift of the response fields.
In this appendix, we show that the Yaglom relation can also be inferred from the symmetries
 of the path integral.

Let us consider the following spacetime-dependent field transformation 
$$\bar{\theta}\left(t,\vx\right)\rightarrow\bar{\theta}\left(t,\vx\right)+\varepsilon\left(t,\vx\right),\quad
\bar{p}\left(t,\vx\right)\rightarrow \bar{p}\left(t,\vx\right)+\varepsilon\left(t,\vx\right)\theta\left(t,\vx\right)$$
in the action functional (\ref{eq:action-NS_plus_scalar}). The variation of
the action reads
\begin{eqnarray*}
&&
\delta S = \int_{t\vx}\varepsilon\left(t,\vx\right)
\Bigr( \partial_{t}\theta\left(t,\vx\right)+\partial_{i}\left(v^{i}\left(t,\vx\right)\theta\left(t,\vx\right)\right) \\
&& \qquad\qquad\qquad\qquad\qquad\qquad
-\frac{\kappa}{2}\partial^{2}\theta\left(t,\vx\right)-M_{k}\bar{\theta}\left(t,\vx\right) \Bigr)\,.
\end{eqnarray*}
Hence, by performing this change of variables in the path integral, and writing that it must leave it unchanged,
one obtains
\begin{eqnarray}
&&
\Biggr\langle\partial_{t}\theta\left(t,\vx\right)+\partial_{i}\left(v^{i}\left(t,\vx\right)\theta\left(t,\vx\right)\right) 
\qquad\qquad\qquad\qquad\qquad \label{eq:fully-gauged-shift-WI-1} \\
&& \qquad \qquad 
-\frac{\kappa}{2}\partial^{2}\theta\left(t,\vx\right)-\int_{\vy} M_{k}(\vx-\vy) \bar{\theta}\left(t,\vx\right)\Biggr\rangle_{J_{\bar{\theta}}=J_{\bar{p}}=0} = 0\,,\nonumber
\end{eqnarray}
where the subscript indicates that the sources associated to $\bar{\theta}$
and $\bar{p}$ are taken to zero (while  the other
ones are kept generic). We now want to express Eq.~(\ref{eq:fully-gauged-shift-WI-1})
in terms of functional derivatives of the generating functional $W\left[J\right]$.
However, the second term in (\ref{eq:fully-gauged-shift-WI-1})
is nonlinear in the fields,  it represents an insertion of a composite
operator. Therefore, we introduce a further source conjugate to the
composite operator $v^{i}\left(t,\vx\right)\theta\left(t,\vx\right)$
by adding the term $\int_{t\vx}L_{i}\left(t,\vx\right)v^{i}\left(t,\vx\right)\theta\left(t,\vx\right)$ to the action.
One then can rewrite equation (\ref{eq:fully-gauged-shift-WI-1})
as
\begin{eqnarray*}
&&
\Big(\partial_{t_{x}}-\frac{\kappa}{2}\partial_x^{2} \Big)\frac{\delta W}{\delta J_{\theta}\left(t_{x},\vx\right)}+
\partial_x^{i}\frac{\delta W}{\delta L_{i}\left(t_{x},\vx\right)} \\
&& \qquad\qquad\qquad\qquad\qquad
-\int_{\vy} M_{k}\left(\vx-\vy\right)\frac{\delta W}{\delta J_{\bar{\theta}}\left(t_{x},\vy\right)} = 0\,,
\end{eqnarray*}
where we have set $J_{\bar{p}}=J_{\bar{\theta}}=0$ in $W\left[J\right]$. 

We take a further differentiation with respect to $J_{\theta}\left(t_{y},y\right)$
and obtain
\begin{eqnarray*}
&& 
\Big(\partial_{t_{x}} -\frac{\kappa}{2}\partial_x^{2}\Big)\frac{\delta W}{\delta J_{\theta}\left(t_{x},\vx\right)\delta J_{\theta}\left(t_{y},\vy\right)}
+\partial_x^{i}\frac{\delta W}{\delta L_{i}\left(t_{x},\vx\right)\delta J_{\theta}\left(t_{y},\vy\right)} \\
&&\qquad\qquad\qquad
-\int_{\vz}M_{k}\left(\vx-\vz\right)\frac{\delta W}{\delta J_{\bar{\theta}}\left(t_{x},\vz\right)\delta J_{\theta}\left(t_{y},\vy\right)} 
 \,=\,0\,  ,
\end{eqnarray*}
that is, explicitly writing the corresponding connected correlation functions
\begin{eqnarray*}
&&
\Big(\partial_{t_{x}}- \frac{\kappa}{2}\partial_{x}^{2}\Big)\Bigr\langle\theta\left(t_{x},\vx\right)\theta\left(t_{y},\vy\right)\Bigr\rangle+\partial_x^{i}\Bigr\langle v^{i}\left(t_{x},\vx\right)\theta\left(t_{x},\vx\right)\theta\left(t_{y},\vy\right)\Bigr\rangle \\
&&\qquad\qquad\qquad\qquad\qquad
-\Bigr\langle f\left(t_{x},\vx\right)\theta\left(t_{y},\vy\right)\Bigr\rangle = 0\,.
\end{eqnarray*}
This expression can be symmetrized with respect to $x\leftrightarrow y$ as follows  
\begin{eqnarray}
0&=&
\Big(\partial_{t_{x}}-\frac{\kappa}{2}\partial_{x}^{2} \Big)\Bigr\langle\theta\left(t_{x},\vx\right)\theta\left(t_{y},\vy\right)\Bigr\rangle
-\Bigr\langle f\left(t_{x},\vx\right)\theta\left(t_{y},\vy\right)\Bigr\rangle\nonumber\\
&+&\Big(\partial_{t_{y}} -\frac{\kappa}{2}\partial_{y}^{2}\Big)\Bigr\langle\theta\left(t_{x},\vx\right)\theta\left(t_{y},\vy\right)\Bigr\rangle
-\Bigr\langle f\left(t_{y},\vy\right)\theta\left(t_{x},\vx\right)\Bigr\rangle\label{eq:KarmanHoward-like-relation-for-scalar}  \\
 +&
\partial_{x^{i}}&\Bigr\langle v^{i}\left(t_{x},\vx\right)\theta\left(t_{x},\vx\right)\theta\left(t_{y},\vy\right)\Bigr\rangle 
+\partial_{y^{i}}\Bigr\langle v^{i}\left(t_{y},\vy\right)\theta\left(t_{y},\vy\right)\theta\left(t_{x},\vx\right)\Bigr\rangle\, .
\nonumber
\end{eqnarray}
We now focus on the steady state and take the equal-time limit in
Eq.~(\ref{eq:KarmanHoward-like-relation-for-scalar}). Using incompressibility and translational invariance, 
one can rewrite the cubic terms exploiting the relation
\begin{eqnarray*}
 &  & \frac{1}{2}\frac{\partial}{\partial\left(x-y\right)^{i}}\Bigr\langle
 \left|\theta\left(t,\vx\right)-\theta\left(t,\vy\right)\right|^{2}\left(v_{i}\left(t,\vx\right)-v_{i}\left(t,\vy\right)\right)\Bigr\rangle\\
 & = &  -\Bigr\langle 
\theta\left(t,\vx\right)\partial_{y^{i}}\left(v_{i}\left(t,\vy\right)\theta\left(t,\vy\right)\right)-\left(\vx\leftrightarrow \vy\right) \Bigr\rangle \,.
\end{eqnarray*}
where $\frac{\p}{\p (x-y)^i} = (\frac{\p}{\p_x^i}-\frac{\p}{\p_y^i})/2$.
  Thus one obtains 
\begin{eqnarray*}
0 &=& 
-\frac{\kappa}{2}\partial_{x}^{2}\Bigr\langle\theta\left(t,\vx\right)\theta\left(t,\vy\right)\Bigr\rangle-\frac{\kappa}{2}\partial_{y}^{2}\Bigr\langle\theta\left(t,\vx\right)\theta\left(t,\vy\right)\Bigr\rangle\\
&&
-\Bigr\langle f\left(t,\vx\right)\theta\left(t,\vy\right)\Bigr\rangle-\Bigr\langle f\left(t,\vy\right)\theta\left(t,\vx\right)\Bigr\rangle\\
&&
-\frac{1}{2}\frac{\partial}{\partial\left(x-y\right)^{i}}\Bigr\langle\left|\theta\left(t,\vx\right)-\theta\left(t,\vy\right)\right|^{2}\left(v_{i}\left(t,\vx\right)-v_{i}\left(t,\vy\right)\right)\Bigr\rangle \,.
\end{eqnarray*}
Following the reasoning as in  Ref.~\onlinecite{falkovich_fouxon_oz_2010}, we impose
$\Bigr\langle f\left(t,\vx\right)\theta\left(t,\vy\right)\Bigr\rangle=\epsilon_{\theta}$
in the limit of small $\left|\vx-\vy\right|$. Moreover, by further neglecting the
terms proportional to molecular diffusivity, and for $\left|\vx-\vy\right|\rightarrow0$,
we obtain
\begin{eqnarray*}
-\frac{1}{2}\frac{\partial}{\partial\left(x-y\right)^{i}}\Bigr\langle\left|\theta\left(t,\vx\right)-\theta\left(t,\vy\right)\right|^{2}\left(v_{i}\left(t,\vx\right)-v_{i}\left(t,\vy\right)\right)\Bigr\rangle & = & 2\epsilon_{\theta}\,,
\end{eqnarray*}
which is the Yaglom relation. We hence showed that this relation can in fact be inferred solely  from a symmetry of  the action functional (the invariance under gauged shifts of the response fields).
This relation in turn implies that
\begin{eqnarray*}
\Bigr\langle\left|\theta\left(t,\vx\right)-\theta\left(t,\vy\right)\right|^{2}\left(v_{i}\left(t,\vx\right)-v_{i}\left(t,\vy\right)\right)\Bigr\rangle & = & -4\epsilon_{\theta}\frac{\left(x-y\right)^{i}}{d}\,.
\end{eqnarray*}
Thus, if one assumes K41 scaling for the velocity, one deduces that the scalar field
has the same scaling dimensions,  $\theta\left(t,\vx\right)\sim x^{1/3}$.

\section{Spectrum of the scalar in the dissipative range from Dyson equation}
\label{app:dissipative}

The two-point correlation function (\ref{eq:Gqq-t-p-via-Dyson-fin}) is expressed
in terms of the function $F\left(\vp\right)$, which is a complicated
function given by a wavenumber integral. Since we did not assume
  the scalar to be in the inertial range, 
 Eq.~(\ref{eq:Gqq-t-p-via-Dyson-fin}) also holds in the dissipative range. 
 Let us  show that one may estimate from it the behavior
of $\bar{G}_{\theta\theta}\left(t,\vp\right)$ in the dissipative regime.

If the convective term $\int\bar{\theta}v\cdot\partial\theta$ is
 negligible then one is left with pure diffusion, implying
that the two-point correlation function is given by the bare expression of the propagator
$\bar{G}_{\theta\theta}\left(t,\vp\right)$. When the convective term
is not very efficient, let us assume that the behavior of $\bar{G}_{\theta\theta}\left(t,\vp\right)$
is determined by only the first non-trivial correction
 due to the convection. The computation of $\Sigma_{\theta\bar{\theta}}$
is essentially unaffected, except that one assumes that the molecular diffusivity
term dominates over $\Sigma_{\theta\bar{\theta}}$ in the dissipative
range.
As for $\Sigma_{\bar{\theta}\bar{\theta}}$, the calculation can be
simplified as follows
\begin{eqnarray*}
&&
\Sigma_{\bar{\theta}\bar{\theta}}\left(\vp\right) \\
&& 
\approx  \int_{\omega \vq}\frac{D_{0}}{\left(q^{2}+m^{2}\right)^{\frac{d+\varepsilon}{2}}}\left(p^{2}-
\frac{\left(\vp\cdot \vq\right)^{2}}{q^{2}}\right)\frac{M_{k}\left(\vq+\vp\right)}{\omega^{2}+\left(\frac{\kappa}{2}\left(\vq+\vp\right)^{2}\right)^{2}}\\
&& 
=  \int_{\vq}\frac{D_{0}}{\left(\left(\vq-\vp\right)^{2}+m^{2}\right)^{\frac{d+\varepsilon}{2}}}\left(p^{2}-\frac{\left(p\cdot\left(\vq-\vp\right)\right)^{2}}{\left(q-p\right)^{2}}\right)\frac{M_{k}\left(\vq\right)}{\kappa q^{2}}\,.
\end{eqnarray*}
We note that for large values of $\vq$, the forcing term $M_{k}\left(\vq\right)$
suppresses the corresponding contribution in the integral. Hence, the range of $\vq$
which contributes is about the scale at which $M_{k}\left(\vq\right)$
is peaked. For large $\vp$, {\it i.e.} for $|\vp|$ larger than this scale,
 one has $\vq-\vp\approx-\vp$ for
values of $\vq$ relevant for the integral. By expanding for small $\vq$
all the terms depending in $\vq-\vp$ one obtains
\begin{eqnarray*}
\Sigma_{\bar{\theta}\bar{\theta}}\left(\vp\right) & = & \frac{d-1}{\kappa d}\frac{D_{0}}{\left(p^{2}+m^{2}\right)^{\frac{d+\varepsilon}{2}}}\int_{\vq} M_{k}\left(\vq\right)\,.
\end{eqnarray*}
It follows that for large wavenumbers in the dissipative range,
 $\Sigma_{\bar{\theta}\bar{\theta}}\left(\vp\right)\propto p^{-d-\varepsilon}$.
The temporal dependence of $\bar{G}_{\theta\theta}\left(t,\vp\right)$
is always of the form displayed in (\ref{eq:Gqq-t-p-via-Dyson-fin}), both in the inertial and dissipative range.
Thus, there exists a range where the convection term is perturbative but
it nevertheless affects the spectrum. The spectrum in this range is proportional to
\begin{eqnarray*}
\bar{G}_{\theta\theta}\left(0,\vp\right) & = & \Big(M_{k}\left(\vp\right)+\Sigma_{\bar{\theta}\bar{\theta}}\left(\vp\right)\Big)
\int_{\omega}\frac{1}{\omega^{2}+\left(\frac{\kappa_{{\rm ren}}}{2}p^{2}\right)^{2}}\\
 & \approx & \frac{\Sigma_{\bar{\theta}\bar{\theta}}\left(\vp\right)}{\kappa_{{\rm ren}}p^{2}}\,\propto\,p^{-d-2-\varepsilon}\,.
\end{eqnarray*}\\

%

\begin{thebibliography}{60}%
\makeatletter
\providecommand \@ifxundefined [1]{%
 \@ifx{#1\undefined}
}%
\providecommand \@ifnum [1]{%
 \ifnum #1\expandafter \@firstoftwo
 \else \expandafter \@secondoftwo
 \fi
}%
\providecommand \@ifx [1]{%
 \ifx #1\expandafter \@firstoftwo
 \else \expandafter \@secondoftwo
 \fi
}%
\providecommand \natexlab [1]{#1}%
\providecommand \enquote  [1]{``#1''}%
\providecommand \bibnamefont  [1]{#1}%
\providecommand \bibfnamefont [1]{#1}%
\providecommand \citenamefont [1]{#1}%
\providecommand \href@noop [0]{\@secondoftwo}%
\providecommand \href [0]{\begingroup \@sanitize@url \@href}%
\providecommand \@href[1]{\@@startlink{#1}\@@href}%
\providecommand \@@href[1]{\endgroup#1\@@endlink}%
\providecommand \@sanitize@url [0]{\catcode `\\12\catcode `\$12\catcode
  `\&12\catcode `\#12\catcode `\^12\catcode `\_12\catcode `\%12\relax}%
\providecommand \@@startlink[1]{}%
\providecommand \@@endlink[0]{}%
\providecommand \url  [0]{\begingroup\@sanitize@url \@url }%
\providecommand \@url [1]{\endgroup\@href {#1}{\urlprefix }}%
\providecommand \urlprefix  [0]{URL }%
\providecommand \Eprint [0]{\href }%
\providecommand \doibase [0]{http://dx.doi.org/}%
\providecommand \selectlanguage [0]{\@gobble}%
\providecommand \bibinfo  [0]{\@secondoftwo}%
\providecommand \bibfield  [0]{\@secondoftwo}%
\providecommand \translation [1]{[#1]}%
\providecommand \BibitemOpen [0]{}%
\providecommand \bibitemStop [0]{}%
\providecommand \bibitemNoStop [0]{.\EOS\space}%
\providecommand \EOS [0]{\spacefactor3000\relax}%
\providecommand \BibitemShut  [1]{\csname bibitem#1\endcsname}%
\let\auto@bib@innerbib\@empty
\bibitem [{\citenamefont {Sreenivasan}(2019)}]{Sreenivasan18175}%
  \BibitemOpen
  \bibfield  {author} {\bibinfo {author} {\bibfnamefont {K.~R.}\ \bibnamefont
  {Sreenivasan}},\ }\bibfield  {title} {\enquote {\bibinfo {title} {Turbulent
  mixing: A perspective},}\ }\href {\doibase 10.1073/pnas.1800463115}
  {\bibfield  {journal} {\bibinfo  {journal} {Proceedings of the National
  Academy of Sciences}\ }\textbf {\bibinfo {volume} {116}},\ \bibinfo {pages}
  {18175--18183} (\bibinfo {year} {2019})},\ \Eprint
  {http://arxiv.org/abs/https://www.pnas.org/content/116/37/18175.full.pdf}
  {https://www.pnas.org/content/116/37/18175.full.pdf} \BibitemShut {NoStop}%
\bibitem [{\citenamefont {Obukhov}(1949)}]{Obukhov49}%
  \BibitemOpen
  \bibfield  {author} {\bibinfo {author} {\bibfnamefont {A.~M.}\ \bibnamefont
  {Obukhov}},\ }\bibfield  {title} {\enquote {\bibinfo {title} {Structure of
  the temperature field in turbulent flows},}\ }\href@noop {} {\bibfield
  {journal} {\bibinfo  {journal} {Izv. Akad. Nauk SSSR, Ser. Geogr. Geofiz.}\
  }\textbf {\bibinfo {volume} {13}},\ \bibinfo {pages} {58} (\bibinfo {year}
  {1949})}\BibitemShut {NoStop}%
\bibitem [{\citenamefont {Corrsin}(1951)}]{Corrsin51}%
  \BibitemOpen
  \bibfield  {author} {\bibinfo {author} {\bibfnamefont {S.}~\bibnamefont
  {Corrsin}},\ }\bibfield  {title} {\enquote {\bibinfo {title} {On the spectrum
  of isotropic temperature fluctuations in isotropic turbulence},}\ }\href@noop
  {} {\bibfield  {journal} {\bibinfo  {journal} {J. Appl. Phys.}\ }\textbf
  {\bibinfo {volume} {22}},\ \bibinfo {pages} {469} (\bibinfo {year}
  {1951})}\BibitemShut {NoStop}%
\bibitem [{\citenamefont {{Yaglom}}(1949)}]{Yaglom49}%
  \BibitemOpen
  \bibfield  {author} {\bibinfo {author} {\bibfnamefont {A.~M.}\ \bibnamefont
  {{Yaglom}}},\ }\href@noop {} {\bibfield  {journal} {\bibinfo  {journal}
  {{Dokl. Akad. Nauk SSSR, n. Ser.}}\ }\textbf {\bibinfo {volume} {69}},\
  \bibinfo {pages} {743--746} (\bibinfo {year} {1949})}\BibitemShut {NoStop}%
\bibitem [{\citenamefont {Batchelor}(1959)}]{batchelor_1959}%
  \BibitemOpen
  \bibfield  {author} {\bibinfo {author} {\bibfnamefont {G.~K.}\ \bibnamefont
  {Batchelor}},\ }\bibfield  {title} {\enquote {\bibinfo {title} {Small-scale
  variation of convected quantities like temperature in turbulent fluid part 1.
  general discussion and the case of small conductivity},}\ }\href {\doibase
  10.1017/S002211205900009X} {\bibfield  {journal} {\bibinfo  {journal}
  {Journal of Fluid Mechanics}\ }\textbf {\bibinfo {volume} {5}},\ \bibinfo
  {pages} {113--133} (\bibinfo {year} {1959})}\BibitemShut {NoStop}%
\bibitem [{\citenamefont {Kraichnan}(1968)}]{Kraichnan_PoF_68}%
  \BibitemOpen
  \bibfield  {author} {\bibinfo {author} {\bibfnamefont {R.~H.}\ \bibnamefont
  {Kraichnan}},\ }\bibfield  {title} {\enquote {\bibinfo {title} {Small‐scale
  structure of a scalar field convected by turbulence},}\ }\href {\doibase
  https://doi.org/10.1063/1.1692063} {\bibfield  {journal} {\bibinfo  {journal}
  {The Physics of Fluids}\ }\textbf {\bibinfo {volume} {11}},\ \bibinfo {pages}
  {945} (\bibinfo {year} {1968})}\BibitemShut {NoStop}%
\bibitem [{\citenamefont {Delamotte}(2012)}]{Delamotte:2007pf}%
  \BibitemOpen
  \bibfield  {author} {\bibinfo {author} {\bibfnamefont {B.}~\bibnamefont
  {Delamotte}},\ }\bibfield  {title} {\enquote {\bibinfo {title} {{An
  Introduction to the nonperturbative renormalization group}},}\ }\href
  {\doibase 10.1007/978-3-642-27320-9_2} {\bibfield  {journal} {\bibinfo
  {journal} {Lect. Notes Phys.}\ }\textbf {\bibinfo {volume} {852}},\ \bibinfo
  {pages} {49--132} (\bibinfo {year} {2012})},\ \Eprint
  {http://arxiv.org/abs/cond-mat/0702365} {arXiv:cond-mat/0702365} \BibitemShut
  {NoStop}%
\bibitem [{\citenamefont {Dupuis}\ \emph {et~al.}(2020)\citenamefont {Dupuis},
  \citenamefont {Canet}, \citenamefont {Eichhorn}, \citenamefont {Metzner},
  \citenamefont {Pawlowski}, \citenamefont {Tissier},\ and\ \citenamefont
  {Wschebor}}]{Dupuis:2020fhh}%
  \BibitemOpen
  \bibfield  {author} {\bibinfo {author} {\bibfnamefont {N.}~\bibnamefont
  {Dupuis}}, \bibinfo {author} {\bibfnamefont {L.}~\bibnamefont {Canet}},
  \bibinfo {author} {\bibfnamefont {A.}~\bibnamefont {Eichhorn}}, \bibinfo
  {author} {\bibfnamefont {W.}~\bibnamefont {Metzner}}, \bibinfo {author}
  {\bibfnamefont {J.}~\bibnamefont {Pawlowski}}, \bibinfo {author}
  {\bibfnamefont {M.}~\bibnamefont {Tissier}}, \ and\ \bibinfo {author}
  {\bibfnamefont {N.}~\bibnamefont {Wschebor}},\ }\bibfield  {title} {\enquote
  {\bibinfo {title} {{The nonperturbative functional renormalization group and
  its applications}},}\ }\href@noop {} {\  (\bibinfo {year} {2020})},\ \Eprint
  {http://arxiv.org/abs/2006.04853} {arXiv:2006.04853 [cond-mat.stat-mech]}
  \BibitemShut {NoStop}%
\bibitem [{\citenamefont {Canet}\ \emph {et~al.}(2017)\citenamefont {Canet},
  \citenamefont {Rossetto}, \citenamefont {Wschebor},\ and\ \citenamefont
  {Balarac}}]{Canet:2016nmg}%
  \BibitemOpen
  \bibfield  {author} {\bibinfo {author} {\bibfnamefont {L.}~\bibnamefont
  {Canet}}, \bibinfo {author} {\bibfnamefont {V.}~\bibnamefont {Rossetto}},
  \bibinfo {author} {\bibfnamefont {N.}~\bibnamefont {Wschebor}}, \ and\
  \bibinfo {author} {\bibfnamefont {G.}~\bibnamefont {Balarac}},\ }\bibfield
  {title} {\enquote {\bibinfo {title} {{Spatiotemporal velocity-velocity
  correlation function in fully developed turbulence}},}\ }\href {\doibase
  10.1103/PhysRevE.95.023107} {\bibfield  {journal} {\bibinfo  {journal} {Phys.
  Rev. E}\ }\textbf {\bibinfo {volume} {95}},\ \bibinfo {pages} {023107}
  (\bibinfo {year} {2017})},\ \Eprint {http://arxiv.org/abs/1607.03098}
  {arXiv:1607.03098 [physics.flu-dyn]} \BibitemShut {NoStop}%
\bibitem [{\citenamefont {Tarpin}, \citenamefont {Canet},\ and\ \citenamefont
  {Wschebor}(2018)}]{Tarpin:2017uzn}%
  \BibitemOpen
  \bibfield  {author} {\bibinfo {author} {\bibfnamefont {M.}~\bibnamefont
  {Tarpin}}, \bibinfo {author} {\bibfnamefont {L.}~\bibnamefont {Canet}}, \
  and\ \bibinfo {author} {\bibfnamefont {N.}~\bibnamefont {Wschebor}},\
  }\bibfield  {title} {\enquote {\bibinfo {title} {{Breaking of scale
  invariance in the time dependence of correlation functions in isotropic and
  homogeneous turbulence}},}\ }\href {\doibase 10.1063/1.5020022} {\bibfield
  {journal} {\bibinfo  {journal} {Phys. Fluids}\ }\textbf {\bibinfo {volume}
  {30}},\ \bibinfo {pages} {055102} (\bibinfo {year} {2018})},\ \Eprint
  {http://arxiv.org/abs/1707.06809} {arXiv:1707.06809 [cond-mat.stat-mech]}
  \BibitemShut {NoStop}%
\bibitem [{\citenamefont {Gorbunova}\ \emph {et~al.}(2020)\citenamefont
  {Gorbunova}, \citenamefont {Balarac}, \citenamefont {Bourgoin}, \citenamefont
  {Canet}, \citenamefont {Mordant},\ and\ \citenamefont
  {Rossetto}}]{Gorbunova:2020uxl}%
  \BibitemOpen
  \bibfield  {author} {\bibinfo {author} {\bibfnamefont {A.}~\bibnamefont
  {Gorbunova}}, \bibinfo {author} {\bibfnamefont {G.}~\bibnamefont {Balarac}},
  \bibinfo {author} {\bibfnamefont {M.}~\bibnamefont {Bourgoin}}, \bibinfo
  {author} {\bibfnamefont {L.}~\bibnamefont {Canet}}, \bibinfo {author}
  {\bibfnamefont {N.}~\bibnamefont {Mordant}}, \ and\ \bibinfo {author}
  {\bibfnamefont {V.}~\bibnamefont {Rossetto}},\ }\bibfield  {title} {\enquote
  {\bibinfo {title} {{Analysis of the dissipative range of the energy spectrum
  in grid turbulence and in direct numerical simulations}},}\ }\href {\doibase
  10.1103/PhysRevFluids.5.044604} {\bibfield  {journal} {\bibinfo  {journal}
  {Phys. Rev. Fluids}\ }\textbf {\bibinfo {volume} {5}},\ \bibinfo {pages}
  {044604} (\bibinfo {year} {2020})},\ \Eprint
  {http://arxiv.org/abs/1909.02907} {arXiv:1909.02907 [physics.flu-dyn]}
  \BibitemShut {NoStop}%
\bibitem [{\citenamefont {Gorbunova}\ \emph
  {et~al.}({\natexlab{a}})\citenamefont {Gorbunova}, \citenamefont {Balarac},
  \citenamefont {Canet}, \citenamefont {Eyink},\ and\ \citenamefont
  {Rossetto}}]{Gorbunova:2021a}%
  \BibitemOpen
  \bibfield  {author} {\bibinfo {author} {\bibfnamefont {A.}~\bibnamefont
  {Gorbunova}}, \bibinfo {author} {\bibfnamefont {G.}~\bibnamefont {Balarac}},
  \bibinfo {author} {\bibfnamefont {L.}~\bibnamefont {Canet}}, \bibinfo
  {author} {\bibfnamefont {G.}~\bibnamefont {Eyink}}, \ and\ \bibinfo {author}
  {\bibfnamefont {V.}~\bibnamefont {Rossetto}},\ }\bibfield  {title} {\enquote
  {\bibinfo {title} {{Spatio-temporal correlations in 3D homogeneous isotropic
  turbulence}},}\ }\href@noop {} {\  ({\natexlab{a}})},\ \Eprint
  {http://arxiv.org/abs/2102.02858} {arXiv:2102.02858 [physics.flu-dyn]}
  \BibitemShut {NoStop}%
\bibitem [{\citenamefont {Tarpin}\ \emph {et~al.}(2019)\citenamefont {Tarpin},
  \citenamefont {Canet}, \citenamefont {Pagani},\ and\ \citenamefont
  {Wschebor}}]{Tarpin:2018yvs}%
  \BibitemOpen
  \bibfield  {author} {\bibinfo {author} {\bibfnamefont {M.}~\bibnamefont
  {Tarpin}}, \bibinfo {author} {\bibfnamefont {L.}~\bibnamefont {Canet}},
  \bibinfo {author} {\bibfnamefont {C.}~\bibnamefont {Pagani}}, \ and\ \bibinfo
  {author} {\bibfnamefont {N.}~\bibnamefont {Wschebor}},\ }\bibfield  {title}
  {\enquote {\bibinfo {title} {{Stationary, isotropic and homogeneous
  two-dimensional turbulence: a first non-perturbative renormalization group
  approach}},}\ }\href {\doibase 10.1088/1751-8121/aaf3f0} {\bibfield
  {journal} {\bibinfo  {journal} {J. Phys. A}\ }\textbf {\bibinfo {volume}
  {52}},\ \bibinfo {pages} {085501} (\bibinfo {year} {2019})},\ \Eprint
  {http://arxiv.org/abs/1809.00909} {arXiv:1809.00909 [cond-mat.stat-mech]}
  \BibitemShut {NoStop}%
\bibitem [{\citenamefont {Kraichnan}(1994)}]{Kraichnan94prl}%
  \BibitemOpen
  \bibfield  {author} {\bibinfo {author} {\bibfnamefont {R.~H.}\ \bibnamefont
  {Kraichnan}},\ }\bibfield  {title} {\enquote {\bibinfo {title} {Anomalous
  scaling of a randomly advected passive scalar},}\ }\href {\doibase
  10.1103/PhysRevLett.72.1016} {\bibfield  {journal} {\bibinfo  {journal}
  {Phys. Rev. Lett.}\ }\textbf {\bibinfo {volume} {72}},\ \bibinfo {pages}
  {1016--1019} (\bibinfo {year} {1994})}\BibitemShut {NoStop}%
\bibitem [{\citenamefont {Chertkov}\ \emph {et~al.}(1995)\citenamefont
  {Chertkov}, \citenamefont {Falkovich}, \citenamefont {Kolokolov},\ and\
  \citenamefont {Lebedev}}]{ChertkovETal1995}%
  \BibitemOpen
  \bibfield  {author} {\bibinfo {author} {\bibfnamefont {M.}~\bibnamefont
  {Chertkov}}, \bibinfo {author} {\bibfnamefont {G.}~\bibnamefont {Falkovich}},
  \bibinfo {author} {\bibfnamefont {I.}~\bibnamefont {Kolokolov}}, \ and\
  \bibinfo {author} {\bibfnamefont {V.}~\bibnamefont {Lebedev}},\ }\bibfield
  {title} {\enquote {\bibinfo {title} {Normal and anomalous scaling of the
  fourth-order correlation function of a randomly advected passive scalar},}\
  }\href {\doibase 10.1103/PhysRevE.52.4924} {\bibfield  {journal} {\bibinfo
  {journal} {Phys. Rev. E}\ }\textbf {\bibinfo {volume} {52}},\ \bibinfo
  {pages} {4924--4941} (\bibinfo {year} {1995})}\BibitemShut {NoStop}%
\bibitem [{\citenamefont {Gawedzki}\ and\ \citenamefont
  {Kupiainen}(1995)}]{Gawedzki:1995zz}%
  \BibitemOpen
  \bibfield  {author} {\bibinfo {author} {\bibfnamefont {K.}~\bibnamefont
  {Gawedzki}}\ and\ \bibinfo {author} {\bibfnamefont {A.}~\bibnamefont
  {Kupiainen}},\ }\bibfield  {title} {\enquote {\bibinfo {title} {{Anomalous
  Scaling of the Passive Scalar}},}\ }\href {\doibase
  10.1103/PhysRevLett.75.3834} {\bibfield  {journal} {\bibinfo  {journal}
  {Phys. Rev. Lett.}\ }\textbf {\bibinfo {volume} {75}},\ \bibinfo {pages}
  {3834--3837} (\bibinfo {year} {1995})},\ \Eprint
  {http://arxiv.org/abs/chao-dyn/9506010} {arXiv:chao-dyn/9506010} \BibitemShut
  {NoStop}%
\bibitem [{\citenamefont {Chertkov}\ and\ \citenamefont
  {Falkovich}(1996)}]{ChertkovFalkovich1996}%
  \BibitemOpen
  \bibfield  {author} {\bibinfo {author} {\bibfnamefont {M.}~\bibnamefont
  {Chertkov}}\ and\ \bibinfo {author} {\bibfnamefont {G.}~\bibnamefont
  {Falkovich}},\ }\bibfield  {title} {\enquote {\bibinfo {title} {Anomalous
  scaling exponents of a white-advected passive scalar},}\ }\href {\doibase
  10.1103/PhysRevLett.76.2706} {\bibfield  {journal} {\bibinfo  {journal}
  {Phys. Rev. Lett.}\ }\textbf {\bibinfo {volume} {76}},\ \bibinfo {pages}
  {2706--2709} (\bibinfo {year} {1996})}\BibitemShut {NoStop}%
\bibitem [{\citenamefont {Bernard}, \citenamefont {Gawedzki},\ and\
  \citenamefont {Kupiainen}(1996)}]{Bernard:1996um}%
  \BibitemOpen
  \bibfield  {author} {\bibinfo {author} {\bibfnamefont {D.}~\bibnamefont
  {Bernard}}, \bibinfo {author} {\bibfnamefont {K.}~\bibnamefont {Gawedzki}}, \
  and\ \bibinfo {author} {\bibfnamefont {A.}~\bibnamefont {Kupiainen}},\
  }\bibfield  {title} {\enquote {\bibinfo {title} {{Anomalous scaling in the N
  point functions of passive scalar}},}\ }\href {\doibase
  10.1103/PhysRevE.54.2564} {\bibfield  {journal} {\bibinfo  {journal} {Phys.
  Rev. E}\ }\textbf {\bibinfo {volume} {54}},\ \bibinfo {pages} {2564}
  (\bibinfo {year} {1996})},\ \Eprint {http://arxiv.org/abs/chao-dyn/9601018}
  {arXiv:chao-dyn/9601018} \BibitemShut {NoStop}%
\bibitem [{\citenamefont {Bernard}, \citenamefont {Gawedzki},\ and\
  \citenamefont {Kupiainen}(1998)}]{BernardGK1998}%
  \BibitemOpen
  \bibfield  {author} {\bibinfo {author} {\bibfnamefont {D.}~\bibnamefont
  {Bernard}}, \bibinfo {author} {\bibfnamefont {K.}~\bibnamefont {Gawedzki}}, \
  and\ \bibinfo {author} {\bibfnamefont {A.}~\bibnamefont {Kupiainen}},\
  }\bibfield  {title} {\enquote {\bibinfo {title} {{Slow Modes in Passive
  Advection}},}\ }\href {\doibase https://doi.org/10.1023/A:1023212600779}
  {\bibfield  {journal} {\bibinfo  {journal} {Journal of Statistical Physics}\
  }\textbf {\bibinfo {volume} {90}},\ \bibinfo {pages} {519} (\bibinfo {year}
  {1998})},\ \Eprint {http://arxiv.org/abs/cond-mat/9706035}
  {arXiv:cond-mat/9706035} \BibitemShut {NoStop}%
\bibitem [{\citenamefont {Adzhemyan}, \citenamefont {Antonov},\ and\
  \citenamefont {Vasil'ev}(1998)}]{PhysRevE.58.1823}%
  \BibitemOpen
  \bibfield  {author} {\bibinfo {author} {\bibfnamefont {L.~T.}\ \bibnamefont
  {Adzhemyan}}, \bibinfo {author} {\bibfnamefont {N.~V.}\ \bibnamefont
  {Antonov}}, \ and\ \bibinfo {author} {\bibfnamefont {A.~N.}\ \bibnamefont
  {Vasil'ev}},\ }\bibfield  {title} {\enquote {\bibinfo {title}
  {Renormalization group, operator product expansion, and anomalous scaling in
  a model of advected passive scalar},}\ }\href {\doibase
  10.1103/PhysRevE.58.1823} {\bibfield  {journal} {\bibinfo  {journal} {Phys.
  Rev. E}\ }\textbf {\bibinfo {volume} {58}},\ \bibinfo {pages} {1823--1835}
  (\bibinfo {year} {1998})}\BibitemShut {NoStop}%
\bibitem [{\citenamefont {Adzhemyan}\ \emph {et~al.}(2001)\citenamefont
  {Adzhemyan}, \citenamefont {Antonov}, \citenamefont {Barinov}, \citenamefont
  {Kabrits},\ and\ \citenamefont {Vasil'ev}}]{PhysRevE.63.025303}%
  \BibitemOpen
  \bibfield  {author} {\bibinfo {author} {\bibfnamefont {L.~T.}\ \bibnamefont
  {Adzhemyan}}, \bibinfo {author} {\bibfnamefont {N.~V.}\ \bibnamefont
  {Antonov}}, \bibinfo {author} {\bibfnamefont {V.~A.}\ \bibnamefont
  {Barinov}}, \bibinfo {author} {\bibfnamefont {Y.~S.}\ \bibnamefont
  {Kabrits}}, \ and\ \bibinfo {author} {\bibfnamefont {A.~N.}\ \bibnamefont
  {Vasil'ev}},\ }\bibfield  {title} {\enquote {\bibinfo {title} {Anomalous
  exponents to order ${\ensuremath{\varepsilon}}^{3}$ in the rapid-change model
  of passive scalar advection},}\ }\href {\doibase 10.1103/PhysRevE.63.025303}
  {\bibfield  {journal} {\bibinfo  {journal} {Phys. Rev. E}\ }\textbf {\bibinfo
  {volume} {63}},\ \bibinfo {pages} {025303} (\bibinfo {year}
  {2001})}\BibitemShut {NoStop}%
\bibitem [{\citenamefont {Adzhemyan}\ and\ \citenamefont
  {Antonov}(1998)}]{PhysRevE.58.7381}%
  \BibitemOpen
  \bibfield  {author} {\bibinfo {author} {\bibfnamefont {L.~T.}\ \bibnamefont
  {Adzhemyan}}\ and\ \bibinfo {author} {\bibfnamefont {N.~V.}\ \bibnamefont
  {Antonov}},\ }\bibfield  {title} {\enquote {\bibinfo {title} {Renormalization
  group and anomalous scaling in a simple model of passive scalar advection in
  compressible flow},}\ }\href {\doibase 10.1103/PhysRevE.58.7381} {\bibfield
  {journal} {\bibinfo  {journal} {Phys. Rev. E}\ }\textbf {\bibinfo {volume}
  {58}},\ \bibinfo {pages} {7381--7396} (\bibinfo {year} {1998})}\BibitemShut
  {NoStop}%
\bibitem [{\citenamefont {Kupiainen}\ and\ \citenamefont
  {Muratore-Ginanneschi}(2007)}]{Kupiainen:2006em}%
  \BibitemOpen
  \bibfield  {author} {\bibinfo {author} {\bibfnamefont {A.}~\bibnamefont
  {Kupiainen}}\ and\ \bibinfo {author} {\bibfnamefont {P.}~\bibnamefont
  {Muratore-Ginanneschi}},\ }\bibfield  {title} {\enquote {\bibinfo {title}
  {{Scaling, renormalization and statistical conservation laws in the Kraichnan
  model of turbulent advection}},}\ }\href {\doibase 10.1007/s10955-006-9205-9}
  {\bibfield  {journal} {\bibinfo  {journal} {J. Statist. Phys.}\ }\textbf
  {\bibinfo {volume} {126}},\ \bibinfo {pages} {669} (\bibinfo {year}
  {2007})},\ \Eprint {http://arxiv.org/abs/nlin/0603031} {arXiv:nlin/0603031}
  \BibitemShut {NoStop}%
\bibitem [{\citenamefont {Pagani}(2015)}]{Pagani:2015hna}%
  \BibitemOpen
  \bibfield  {author} {\bibinfo {author} {\bibfnamefont {C.}~\bibnamefont
  {Pagani}},\ }\bibfield  {title} {\enquote {\bibinfo {title} {{Functional
  Renormalization Group approach to the Kraichnan model}},}\ }\href {\doibase
  10.1103/PhysRevE.92.033016} {\bibfield  {journal} {\bibinfo  {journal} {Phys.
  Rev. E}\ }\textbf {\bibinfo {volume} {92}},\ \bibinfo {pages} {033016}
  (\bibinfo {year} {2015})},\ \bibinfo {note} {[Addendum: Phys.Rev.E 97, 049902
  (2018)]},\ \Eprint {http://arxiv.org/abs/1505.01293} {arXiv:1505.01293
  [cond-mat.stat-mech]} \BibitemShut {NoStop}%
\bibitem [{\citenamefont {Martin}, \citenamefont {Siggia},\ and\ \citenamefont
  {Rose}(1973)}]{Martin73}%
  \BibitemOpen
  \bibfield  {author} {\bibinfo {author} {\bibfnamefont {P.~C.}\ \bibnamefont
  {Martin}}, \bibinfo {author} {\bibfnamefont {E.~D.}\ \bibnamefont {Siggia}},
  \ and\ \bibinfo {author} {\bibfnamefont {H.~A.}\ \bibnamefont {Rose}},\
  }\bibfield  {title} {\enquote {\bibinfo {title} {Statistical dynamics of
  classical systems},}\ }\href {\doibase 10.1103/PhysRevA.8.423} {\bibfield
  {journal} {\bibinfo  {journal} {Phys. Rev. A}\ }\textbf {\bibinfo {volume}
  {8}},\ \bibinfo {pages} {423--437} (\bibinfo {year} {1973})}\BibitemShut
  {NoStop}%
\bibitem [{\citenamefont {Janssen}(1976)}]{Janssen76}%
  \BibitemOpen
  \bibfield  {author} {\bibinfo {author} {\bibfnamefont {H.-K.}\ \bibnamefont
  {Janssen}},\ }\bibfield  {title} {\enquote {\bibinfo {title} {On a lagrangean
  for classical field dynamics and {R}enormalization {G}roup calculations of
  dynamical critical properties},}\ }\href {\doibase 10.1007/BF01316547}
  {\bibfield  {journal} {\bibinfo  {journal} {Z. Phys. B}\ }\textbf {\bibinfo
  {volume} {23}},\ \bibinfo {pages} {377--380} (\bibinfo {year}
  {1976})}\BibitemShut {NoStop}%
\bibitem [{\citenamefont {de~Dominicis}(1976)}]{Dominicis76}%
  \BibitemOpen
  \bibfield  {author} {\bibinfo {author} {\bibfnamefont {C.}~\bibnamefont
  {de~Dominicis}},\ }\bibfield  {title} {\enquote {\bibinfo {title} {Techniques
  de renormalisation de la th\'eorie des champs et dynamique des ph\'enom\`enes
  critiques},}\ }\href {\doibase 10.1051/jphyscol:1976138} {\bibfield
  {journal} {\bibinfo  {journal} {J. Phys. (Paris) Colloq.}\ }\textbf {\bibinfo
  {volume} {37}},\ \bibinfo {pages} {247--253} (\bibinfo {year}
  {1976})}\BibitemShut {NoStop}%
\bibitem [{\citenamefont {DeDominicis}\ and\ \citenamefont
  {Martin}(1979)}]{Dominicis79}%
  \BibitemOpen
  \bibfield  {author} {\bibinfo {author} {\bibfnamefont {C.}~\bibnamefont
  {DeDominicis}}\ and\ \bibinfo {author} {\bibfnamefont {P.~C.}\ \bibnamefont
  {Martin}},\ }\bibfield  {title} {\enquote {\bibinfo {title} {Energy spectra
  of certain randomly-stirred fluids},}\ }\href {\doibase
  10.1103/PhysRevA.19.419} {\bibfield  {journal} {\bibinfo  {journal} {Phys.
  Rev. A}\ }\textbf {\bibinfo {volume} {19}},\ \bibinfo {pages} {419--422}
  (\bibinfo {year} {1979})}\BibitemShut {NoStop}%
\bibitem [{\citenamefont {Fournier}\ and\ \citenamefont
  {Frisch}(1983)}]{Fournier83}%
  \BibitemOpen
  \bibfield  {author} {\bibinfo {author} {\bibfnamefont {J.~D.}\ \bibnamefont
  {Fournier}}\ and\ \bibinfo {author} {\bibfnamefont {U.}~\bibnamefont
  {Frisch}},\ }\bibfield  {title} {\enquote {\bibinfo {title} {Remarks on the
  renormalization group in statistical fluid dynamics},}\ }\href {\doibase
  10.1103/PhysRevA.28.1000} {\bibfield  {journal} {\bibinfo  {journal} {Phys.
  Rev. A}\ }\textbf {\bibinfo {volume} {28}},\ \bibinfo {pages} {1000--1002}
  (\bibinfo {year} {1983})}\BibitemShut {NoStop}%
\bibitem [{\citenamefont {Yakhot}\ and\ \citenamefont
  {Orszag}(1986)}]{Yakhot1986}%
  \BibitemOpen
  \bibfield  {author} {\bibinfo {author} {\bibfnamefont {V.}~\bibnamefont
  {Yakhot}}\ and\ \bibinfo {author} {\bibfnamefont {S.~A.}\ \bibnamefont
  {Orszag}},\ }\bibfield  {title} {\enquote {\bibinfo {title} {Renormalization
  group analysis of turbulence. i. basic theory},}\ }\href {\doibase
  10.1007/BF01061452} {\bibfield  {journal} {\bibinfo  {journal} {Journal of
  Scientific Computing}\ }\textbf {\bibinfo {volume} {1}},\ \bibinfo {pages}
  {3--51} (\bibinfo {year} {1986})}\BibitemShut {NoStop}%
\bibitem [{\citenamefont {Canuto}\ and\ \citenamefont {S.}(1996)}]{Canuto96}%
  \BibitemOpen
  \bibfield  {author} {\bibinfo {author} {\bibfnamefont {V.~M.}\ \bibnamefont
  {Canuto}}\ and\ \bibinfo {author} {\bibfnamefont {D.~M.}\ \bibnamefont
  {S.}},\ }\bibfield  {title} {\enquote {\bibinfo {title} {A dynamical model
  for turbulence. i. general formalism},}\ }\href
  {https://doi.org/10.1063/1.868842} {\bibfield  {journal} {\bibinfo  {journal}
  {Physics of Fluids}\ }\textbf {\bibinfo {volume} {8}},\ \bibinfo {pages}
  {571} (\bibinfo {year} {1996})}\BibitemShut {NoStop}%
\bibitem [{\citenamefont {L.}\ \emph {et~al.}(17)\citenamefont {L.},
  \citenamefont {N.}, \citenamefont {M.},\ and\ \citenamefont
  {A.}}]{Adzhemyan03}%
  \BibitemOpen
  \bibfield  {author} {\bibinfo {author} {\bibfnamefont {A.}~\bibnamefont
  {L.}}, \bibinfo {author} {\bibfnamefont {A.}~\bibnamefont {N.}}, \bibinfo
  {author} {\bibfnamefont {K.}~\bibnamefont {M.}}, \ and\ \bibinfo {author}
  {\bibfnamefont {V.}~\bibnamefont {A.}},\ }\bibfield  {title} {\enquote
  {\bibinfo {title} {Renormalization-group approach to the stochastic
  navier–stokes equation: Two-loop approximation},}\ }\href
  {https://doi.org/10.1142/S0217979203018193} {\bibfield  {journal} {\bibinfo
  {journal} {Int. J. Mod. Phys.}\ }\textbf {\bibinfo {volume} {8}},\ \bibinfo
  {pages} {2137--2170} (\bibinfo {year} {17})}\BibitemShut {NoStop}%
\bibitem [{\citenamefont {L.}, \citenamefont {N.},\ and\ \citenamefont
  {A.}(1999)}]{Adzhemyan1999}%
  \BibitemOpen
  \bibfield  {author} {\bibinfo {author} {\bibfnamefont {A.}~\bibnamefont
  {L.}}, \bibinfo {author} {\bibfnamefont {A.}~\bibnamefont {N.}}, \ and\
  \bibinfo {author} {\bibfnamefont {V.}~\bibnamefont {A.}},\ }\href@noop {}
  {\emph {\bibinfo {title} {The Field Theoretic Renormalization Group in Fully
  Developed Turbulence}}}\ (\bibinfo  {publisher} {Gordon and Breach},\
  \bibinfo {address} {London},\ \bibinfo {year} {1999})\BibitemShut {NoStop}%
\bibitem [{\citenamefont {Wetterich}(1993)}]{Wetterich:1992yh}%
  \BibitemOpen
  \bibfield  {author} {\bibinfo {author} {\bibfnamefont {C.}~\bibnamefont
  {Wetterich}},\ }\bibfield  {title} {\enquote {\bibinfo {title} {{Exact
  evolution equation for the effective potential}},}\ }\href {\doibase
  10.1016/0370-2693(93)90726-X} {\bibfield  {journal} {\bibinfo  {journal}
  {Phys. Lett. B}\ }\textbf {\bibinfo {volume} {301}},\ \bibinfo {pages}
  {90--94} (\bibinfo {year} {1993})},\ \Eprint
  {http://arxiv.org/abs/1710.05815} {arXiv:1710.05815 [hep-th]} \BibitemShut
  {NoStop}%
\bibitem [{\citenamefont {Reuter}\ and\ \citenamefont
  {Wetterich}(1994{\natexlab{a}})}]{Reuter:1993kw}%
  \BibitemOpen
  \bibfield  {author} {\bibinfo {author} {\bibfnamefont {M.}~\bibnamefont
  {Reuter}}\ and\ \bibinfo {author} {\bibfnamefont {C.}~\bibnamefont
  {Wetterich}},\ }\bibfield  {title} {\enquote {\bibinfo {title} {{Effective
  average action for gauge theories and exact evolution equations}},}\ }\href
  {\doibase 10.1016/0550-3213(94)90543-6} {\bibfield  {journal} {\bibinfo
  {journal} {Nucl. Phys. B}\ }\textbf {\bibinfo {volume} {417}},\ \bibinfo
  {pages} {181--214} (\bibinfo {year} {1994}{\natexlab{a}})}\BibitemShut
  {NoStop}%
\bibitem [{\citenamefont {Reuter}\ and\ \citenamefont
  {Wetterich}(1994{\natexlab{b}})}]{Reuter:1994sg}%
  \BibitemOpen
  \bibfield  {author} {\bibinfo {author} {\bibfnamefont {M.}~\bibnamefont
  {Reuter}}\ and\ \bibinfo {author} {\bibfnamefont {C.}~\bibnamefont
  {Wetterich}},\ }\bibfield  {title} {\enquote {\bibinfo {title} {{Exact
  evolution equation for scalar electrodynamics}},}\ }\href {\doibase
  10.1016/0550-3213(94)90278-X} {\bibfield  {journal} {\bibinfo  {journal}
  {Nucl. Phys. B}\ }\textbf {\bibinfo {volume} {427}},\ \bibinfo {pages}
  {291--324} (\bibinfo {year} {1994}{\natexlab{b}})}\BibitemShut {NoStop}%
\bibitem [{\citenamefont {Ellwanger}(1993)}]{Ellwanger:1993mw}%
  \BibitemOpen
  \bibfield  {author} {\bibinfo {author} {\bibfnamefont {U.}~\bibnamefont
  {Ellwanger}},\ }\bibfield  {title} {\enquote {\bibinfo {title} {{FLow
  equations for N point functions and bound states}},}\ }\href {\doibase
  10.1007/BF01555911} {\ ,\ \bibinfo {pages} {206--211} (\bibinfo {year}
  {1993})},\ \Eprint {http://arxiv.org/abs/hep-ph/9308260}
  {arXiv:hep-ph/9308260} \BibitemShut {NoStop}%
\bibitem [{\citenamefont {Morris}(1994)}]{Morris:1993qb}%
  \BibitemOpen
  \bibfield  {author} {\bibinfo {author} {\bibfnamefont {T.~R.}\ \bibnamefont
  {Morris}},\ }\bibfield  {title} {\enquote {\bibinfo {title} {{The Exact
  renormalization group and approximate solutions}},}\ }\href {\doibase
  10.1142/S0217751X94000972} {\bibfield  {journal} {\bibinfo  {journal} {Int.
  J. Mod. Phys. A}\ }\textbf {\bibinfo {volume} {9}},\ \bibinfo {pages}
  {2411--2450} (\bibinfo {year} {1994})},\ \Eprint
  {http://arxiv.org/abs/hep-ph/9308265} {arXiv:hep-ph/9308265} \BibitemShut
  {NoStop}%
\bibitem [{\citenamefont {Canet}, \citenamefont {Delamotte},\ and\
  \citenamefont {Wschebor}(2016)}]{Canet:2014dta}%
  \BibitemOpen
  \bibfield  {author} {\bibinfo {author} {\bibfnamefont {L.}~\bibnamefont
  {Canet}}, \bibinfo {author} {\bibfnamefont {B.}~\bibnamefont {Delamotte}}, \
  and\ \bibinfo {author} {\bibfnamefont {N.}~\bibnamefont {Wschebor}},\
  }\bibfield  {title} {\enquote {\bibinfo {title} {{Fully developed isotropic
  turbulence: nonperturbative renormalization group formalism and fixed point
  solution}},}\ }\href {\doibase 10.1103/PhysRevE.93.063101} {\bibfield
  {journal} {\bibinfo  {journal} {Phys. Rev. E}\ }\textbf {\bibinfo {volume}
  {93}},\ \bibinfo {pages} {063101} (\bibinfo {year} {2016})},\ \Eprint
  {http://arxiv.org/abs/1411.7780} {arXiv:1411.7780 [cond-mat.stat-mech]}
  \BibitemShut {NoStop}%
\bibitem [{\citenamefont {Canet}, \citenamefont {Delamotte},\ and\
  \citenamefont {Wschebor}(2015)}]{Canet:2014cta}%
  \BibitemOpen
  \bibfield  {author} {\bibinfo {author} {\bibfnamefont {L.}~\bibnamefont
  {Canet}}, \bibinfo {author} {\bibfnamefont {B.}~\bibnamefont {Delamotte}}, \
  and\ \bibinfo {author} {\bibfnamefont {N.}~\bibnamefont {Wschebor}},\
  }\bibfield  {title} {\enquote {\bibinfo {title} {{Fully developed isotropic
  turbulence: symmetries and exact identities}},}\ }\href {\doibase
  10.1103/PhysRevE.91.053004} {\bibfield  {journal} {\bibinfo  {journal} {Phys.
  Rev. E}\ }\textbf {\bibinfo {volume} {91}},\ \bibinfo {pages} {053004}
  (\bibinfo {year} {2015})},\ \Eprint {http://arxiv.org/abs/1411.7778}
  {arXiv:1411.7778 [cond-mat.stat-mech]} \BibitemShut {NoStop}%
\bibitem [{\citenamefont {Canet}, \citenamefont {Chat\'e},\ and\ \citenamefont
  {Delamotte}(2011)}]{Canet11b}%
  \BibitemOpen
  \bibfield  {author} {\bibinfo {author} {\bibfnamefont {L.}~\bibnamefont
  {Canet}}, \bibinfo {author} {\bibfnamefont {H.}~\bibnamefont {Chat\'e}}, \
  and\ \bibinfo {author} {\bibfnamefont {B.}~\bibnamefont {Delamotte}},\
  }\bibfield  {title} {\enquote {\bibinfo {title} {General framework of the
  non-perturbative {R}enormalization {G}roup for non-equilibrium steady
  states},}\ }\href {http://stacks.iop.org/1751-8121/44/i=49/a=495001}
  {\bibfield  {journal} {\bibinfo  {journal} {J. Phys. A}\ }\textbf {\bibinfo
  {volume} {44}},\ \bibinfo {pages} {495001} (\bibinfo {year}
  {2011})}\BibitemShut {NoStop}%
\bibitem [{Note1()}]{Note1}%
  \BibitemOpen
  \bibinfo {note} {This non-renormalization of the velocity covariance can also
  shown using different arguments, as in the approach of Ref.~\protect
  \rev@citealpnum {PhysRevE.58.1823}.}\BibitemShut {Stop}%
\bibitem [{Note2()}]{Note2}%
  \BibitemOpen
  \bibinfo {note} {Note that this implies that the engineering dimension of the
  constant $D_{0}$ is not the same as its (trivial) scaling
  dimension.}\BibitemShut {Stop}%
\bibitem [{\citenamefont {Blaizot}, \citenamefont {Mendez~Galain},\ and\
  \citenamefont {Wschebor}(2006)}]{Blaizot:2005xy}%
  \BibitemOpen
  \bibfield  {author} {\bibinfo {author} {\bibfnamefont {J.~P.}\ \bibnamefont
  {Blaizot}}, \bibinfo {author} {\bibfnamefont {R.}~\bibnamefont
  {Mendez~Galain}}, \ and\ \bibinfo {author} {\bibfnamefont {N.}~\bibnamefont
  {Wschebor}},\ }\bibfield  {title} {\enquote {\bibinfo {title} {{A New method
  to solve the non perturbative renormalization group equations}},}\ }\href
  {\doibase 10.1016/j.physletb.2005.10.086} {\bibfield  {journal} {\bibinfo
  {journal} {Phys. Lett. B}\ }\textbf {\bibinfo {volume} {632}},\ \bibinfo
  {pages} {571--578} (\bibinfo {year} {2006})},\ \Eprint
  {http://arxiv.org/abs/hep-th/0503103} {arXiv:hep-th/0503103} \BibitemShut
  {NoStop}%
\bibitem [{\citenamefont {Blaizot}, \citenamefont {Mendez-Galain},\ and\
  \citenamefont {Wschebor}(2006{\natexlab{a}})}]{Blaizot:2005wd}%
  \BibitemOpen
  \bibfield  {author} {\bibinfo {author} {\bibfnamefont {J.-P.}\ \bibnamefont
  {Blaizot}}, \bibinfo {author} {\bibfnamefont {R.}~\bibnamefont
  {Mendez-Galain}}, \ and\ \bibinfo {author} {\bibfnamefont {N.}~\bibnamefont
  {Wschebor}},\ }\bibfield  {title} {\enquote {\bibinfo {title} {{Non
  perturbative renormalisation group and momentum dependence of n-point
  functions (I)}},}\ }\href {\doibase 10.1103/PhysRevE.74.051116} {\bibfield
  {journal} {\bibinfo  {journal} {Phys. Rev. E}\ }\textbf {\bibinfo {volume}
  {74}},\ \bibinfo {pages} {051116} (\bibinfo {year} {2006}{\natexlab{a}})},\
  \Eprint {http://arxiv.org/abs/hep-th/0512317} {arXiv:hep-th/0512317}
  \BibitemShut {NoStop}%
\bibitem [{\citenamefont {Blaizot}, \citenamefont {Mendez-Galain},\ and\
  \citenamefont {Wschebor}(2006{\natexlab{b}})}]{Blaizot:2006vr}%
  \BibitemOpen
  \bibfield  {author} {\bibinfo {author} {\bibfnamefont {J.-P.}\ \bibnamefont
  {Blaizot}}, \bibinfo {author} {\bibfnamefont {R.}~\bibnamefont
  {Mendez-Galain}}, \ and\ \bibinfo {author} {\bibfnamefont {N.}~\bibnamefont
  {Wschebor}},\ }\bibfield  {title} {\enquote {\bibinfo {title} {{Non
  perturbative renormalization group and momentum dependence of n-point
  functions. II.}}}\ }\href {\doibase 10.1103/PhysRevE.74.051117} {\bibfield
  {journal} {\bibinfo  {journal} {Phys. Rev. E}\ }\textbf {\bibinfo {volume}
  {74}},\ \bibinfo {pages} {051117} (\bibinfo {year} {2006}{\natexlab{b}})},\
  \Eprint {http://arxiv.org/abs/hep-th/0603163} {arXiv:hep-th/0603163}
  \BibitemShut {NoStop}%
\bibitem [{\citenamefont {Falkovich}, \citenamefont
  {Gaw\ifmmode~\mbox{\c{e}}\else \c{e}\fi{}dzki},\ and\ \citenamefont
  {Vergassola}(2001)}]{RevModPhys.73.913}%
  \BibitemOpen
  \bibfield  {author} {\bibinfo {author} {\bibfnamefont {G.}~\bibnamefont
  {Falkovich}}, \bibinfo {author} {\bibfnamefont {K.}~\bibnamefont
  {Gaw\ifmmode~\mbox{\c{e}}\else \c{e}\fi{}dzki}}, \ and\ \bibinfo {author}
  {\bibfnamefont {M.}~\bibnamefont {Vergassola}},\ }\bibfield  {title}
  {\enquote {\bibinfo {title} {Particles and fields in fluid turbulence},}\
  }\href {\doibase 10.1103/RevModPhys.73.913} {\bibfield  {journal} {\bibinfo
  {journal} {Rev. Mod. Phys.}\ }\textbf {\bibinfo {volume} {73}},\ \bibinfo
  {pages} {913--975} (\bibinfo {year} {2001})}\BibitemShut {NoStop}%
\bibitem [{\citenamefont {Chertkov}, \citenamefont {Falkovich},\ and\
  \citenamefont {Lebedev}(1996)}]{Falkovich1996}%
  \BibitemOpen
  \bibfield  {author} {\bibinfo {author} {\bibfnamefont {M.}~\bibnamefont
  {Chertkov}}, \bibinfo {author} {\bibfnamefont {G.}~\bibnamefont {Falkovich}},
  \ and\ \bibinfo {author} {\bibfnamefont {V.}~\bibnamefont {Lebedev}},\
  }\bibfield  {title} {\enquote {\bibinfo {title} {Nonuniversality of the
  scaling exponents of a passive scalar convected by a random flow},}\ }\href
  {\doibase 10.1103/PhysRevLett.76.3707} {\bibfield  {journal} {\bibinfo
  {journal} {Phys. Rev. Lett.}\ }\textbf {\bibinfo {volume} {76}},\ \bibinfo
  {pages} {3707--3710} (\bibinfo {year} {1996})}\BibitemShut {NoStop}%
\bibitem [{\citenamefont {Eyink}\ and\ \citenamefont
  {Benveniste}(2013)}]{Eyink2013}%
  \BibitemOpen
  \bibfield  {author} {\bibinfo {author} {\bibfnamefont {G.~L.}\ \bibnamefont
  {Eyink}}\ and\ \bibinfo {author} {\bibfnamefont {D.}~\bibnamefont
  {Benveniste}},\ }\bibfield  {title} {\enquote {\bibinfo {title} {Suppression
  of particle dispersion by sweeping effects in synthetic turbulence},}\ }\href
  {\doibase 10.1103/PhysRevE.87.023011} {\bibfield  {journal} {\bibinfo
  {journal} {Phys. Rev. E}\ }\textbf {\bibinfo {volume} {87}},\ \bibinfo
  {pages} {023011} (\bibinfo {year} {2013})}\BibitemShut {NoStop}%
\bibitem [{\citenamefont {Mitra}\ and\ \citenamefont
  {Pandit}(2005)}]{MitraPandit2005}%
  \BibitemOpen
  \bibfield  {author} {\bibinfo {author} {\bibfnamefont {D.}~\bibnamefont
  {Mitra}}\ and\ \bibinfo {author} {\bibfnamefont {R.}~\bibnamefont {Pandit}},\
  }\bibfield  {title} {\enquote {\bibinfo {title} {Dynamics of passive-scalar
  turbulence},}\ }\href {\doibase 10.1103/PhysRevLett.95.144501} {\bibfield
  {journal} {\bibinfo  {journal} {Phys. Rev. Lett.}\ }\textbf {\bibinfo
  {volume} {95}},\ \bibinfo {pages} {144501} (\bibinfo {year}
  {2005})}\BibitemShut {NoStop}%
\bibitem [{Note3()}]{Note3}%
  \BibitemOpen
  \bibinfo {note} {Possible contributions to the flow equation from any
  bare-like three-point vertex actually vanishes by
  incompressibility.}\BibitemShut {Stop}%
\bibitem [{\citenamefont {Gorbunova}\ \emph
  {et~al.}({\natexlab{b}})\citenamefont {Gorbunova}, \citenamefont {C.},
  \citenamefont {Balarac}, \citenamefont {Canet},\ and\ \citenamefont
  {Rossetto}}]{Gorbunova:2021b}%
  \BibitemOpen
  \bibfield  {author} {\bibinfo {author} {\bibfnamefont {A.}~\bibnamefont
  {Gorbunova}}, \bibinfo {author} {\bibfnamefont {P.}~\bibnamefont {C.}},
  \bibinfo {author} {\bibfnamefont {G.}~\bibnamefont {Balarac}}, \bibinfo
  {author} {\bibfnamefont {L.}~\bibnamefont {Canet}}, \ and\ \bibinfo {author}
  {\bibfnamefont {V.}~\bibnamefont {Rossetto}},\ }\bibfield  {title} {\enquote
  {\bibinfo {title} {{Work in preparation}},}\ }\href@noop {} {\
  ({\natexlab{b}})}\BibitemShut {NoStop}%
\bibitem [{Note4()}]{Note4}%
  \BibitemOpen
  \bibinfo {note} {We refer to Ref.~\protect \rev@citealpnum
  {Pagani:2016pad,Pagani:2017tdr,Pagani:2020ejb,Rose:2015bma} for details
  regarding the composite operators within the FRG framework.}\BibitemShut
  {Stop}%
\bibitem [{\citenamefont {Falkovich}, \citenamefont {Fouxon},\ and\
  \citenamefont {Oz}(2010)}]{falkovich_fouxon_oz_2010}%
  \BibitemOpen
  \bibfield  {author} {\bibinfo {author} {\bibfnamefont {G.}~\bibnamefont
  {Falkovich}}, \bibinfo {author} {\bibfnamefont {I.}~\bibnamefont {Fouxon}}, \
  and\ \bibinfo {author} {\bibfnamefont {Y.}~\bibnamefont {Oz}},\ }\bibfield
  {title} {\enquote {\bibinfo {title} {New relations for correlation functions
  in navier–stokes turbulence},}\ }\href {\doibase 10.1017/S0022112009993429}
  {\bibfield  {journal} {\bibinfo  {journal} {Journal of Fluid Mechanics}\
  }\textbf {\bibinfo {volume} {644}},\ \bibinfo {pages} {465--472} (\bibinfo
  {year} {2010})}\BibitemShut {NoStop}%
\bibitem [{\citenamefont {Tomassini}(1997)}]{TOMASSINI1997117}%
  \BibitemOpen
  \bibfield  {author} {\bibinfo {author} {\bibfnamefont {P.}~\bibnamefont
  {Tomassini}},\ }\bibfield  {title} {\enquote {\bibinfo {title} {An exact
  renormalization group analysis of 3d well developed turbulence},}\ }\href
  {\doibase https://doi.org/10.1016/S0370-2693(97)00980-5} {\bibfield
  {journal} {\bibinfo  {journal} {Physics Letters B}\ }\textbf {\bibinfo
  {volume} {411}},\ \bibinfo {pages} {117 -- 126} (\bibinfo {year}
  {1997})}\BibitemShut {NoStop}%
\bibitem [{\citenamefont {Mej\'{\i}a-Monasterio}\ and\ \citenamefont
  {Muratore-Ginanneschi}(2012)}]{Monasterio2012}%
  \BibitemOpen
  \bibfield  {author} {\bibinfo {author} {\bibfnamefont {C.}~\bibnamefont
  {Mej\'{\i}a-Monasterio}}\ and\ \bibinfo {author} {\bibfnamefont
  {P.}~\bibnamefont {Muratore-Ginanneschi}},\ }\bibfield  {title} {\enquote
  {\bibinfo {title} {Nonperturbative renormalization group study of the
  stochastic navier-stokes equation},}\ }\href {\doibase
  10.1103/PhysRevE.86.016315} {\bibfield  {journal} {\bibinfo  {journal} {Phys.
  Rev. E}\ }\textbf {\bibinfo {volume} {86}},\ \bibinfo {pages} {016315}
  (\bibinfo {year} {2012})}\BibitemShut {NoStop}%
\bibitem [{\citenamefont {Pagani}(2016)}]{Pagani:2016pad}%
  \BibitemOpen
  \bibfield  {author} {\bibinfo {author} {\bibfnamefont {C.}~\bibnamefont
  {Pagani}},\ }\bibfield  {title} {\enquote {\bibinfo {title} {{Note on scaling
  arguments in the effective average action formalism}},}\ }\href {\doibase
  10.1103/PhysRevD.94.045001} {\bibfield  {journal} {\bibinfo  {journal} {Phys.
  Rev. D}\ }\textbf {\bibinfo {volume} {94}},\ \bibinfo {pages} {045001}
  (\bibinfo {year} {2016})},\ \Eprint {http://arxiv.org/abs/1603.07250}
  {arXiv:1603.07250 [hep-th]} \BibitemShut {NoStop}%
\bibitem [{\citenamefont {Pagani}\ and\ \citenamefont
  {Sonoda}(2018)}]{Pagani:2017tdr}%
  \BibitemOpen
  \bibfield  {author} {\bibinfo {author} {\bibfnamefont {C.}~\bibnamefont
  {Pagani}}\ and\ \bibinfo {author} {\bibfnamefont {H.}~\bibnamefont
  {Sonoda}},\ }\bibfield  {title} {\enquote {\bibinfo {title} {{Products of
  composite operators in the exact renormalization group formalism}},}\ }\href
  {\doibase 10.1093/ptep/ptx189} {\bibfield  {journal} {\bibinfo  {journal}
  {PTEP}\ }\textbf {\bibinfo {volume} {2018}},\ \bibinfo {pages} {023B02}
  (\bibinfo {year} {2018})},\ \Eprint {http://arxiv.org/abs/1707.09138}
  {arXiv:1707.09138 [hep-th]} \BibitemShut {NoStop}%
\bibitem [{\citenamefont {Pagani}\ and\ \citenamefont
  {Sonoda}(2020)}]{Pagani:2020ejb}%
  \BibitemOpen
  \bibfield  {author} {\bibinfo {author} {\bibfnamefont {C.}~\bibnamefont
  {Pagani}}\ and\ \bibinfo {author} {\bibfnamefont {H.}~\bibnamefont
  {Sonoda}},\ }\bibfield  {title} {\enquote {\bibinfo {title} {{Operator
  product expansion coefficients in the exact renormalization group
  formalism}},}\ }\href {\doibase 10.1103/PhysRevD.101.105007} {\bibfield
  {journal} {\bibinfo  {journal} {Phys. Rev. D}\ }\textbf {\bibinfo {volume}
  {101}},\ \bibinfo {pages} {105007} (\bibinfo {year} {2020})},\ \Eprint
  {http://arxiv.org/abs/2001.07015} {arXiv:2001.07015 [hep-th]} \BibitemShut
  {NoStop}%
\bibitem [{\citenamefont {Rose}, \citenamefont {L\'eonard},\ and\ \citenamefont
  {Dupuis}(2015)}]{Rose:2015bma}%
  \BibitemOpen
  \bibfield  {author} {\bibinfo {author} {\bibfnamefont {F.}~\bibnamefont
  {Rose}}, \bibinfo {author} {\bibfnamefont {F.}~\bibnamefont {L\'eonard}}, \
  and\ \bibinfo {author} {\bibfnamefont {N.}~\bibnamefont {Dupuis}},\
  }\bibfield  {title} {\enquote {\bibinfo {title} {{Higgs amplitude mode in the
  vicinity of a $(2+1)$-dimensional quantum critical point: a nonperturbative
  renormalization-group approach}},}\ }\href {\doibase
  10.1103/PhysRevB.91.224501} {\bibfield  {journal} {\bibinfo  {journal} {Phys.
  Rev. B}\ }\textbf {\bibinfo {volume} {91}},\ \bibinfo {pages} {224501}
  (\bibinfo {year} {2015})},\ \Eprint {http://arxiv.org/abs/1503.08688}
  {arXiv:1503.08688 [cond-mat.quant-gas]} \BibitemShut {NoStop}%
\end{thebibliography}

%

\end{document}